\documentclass[aps,prr,amsmath,amssymb,longbibliography,twocolumn,floatfix]{revtex4-2}

\usepackage{graphicx}
\usepackage{dcolumn}
\usepackage{bm}
\usepackage{amsmath}
\usepackage{amssymb} 
\usepackage{physics}
\usepackage{color}
\usepackage[normalem]{ulem}
\usepackage{xr}
\begin{document}
	
	\title{Physical Constraints on the Rhythmicity of the Biological Clock}
	
	\author{YeongKyu Lee}
	\affiliation{Korea Institute for Advanced Study, Seoul 02455, Korea}
	\author{Changbong Hyeon}
	\email[]{hyeoncb@kias.re.kr}
	\affiliation{Korea Institute for Advanced Study, Seoul 02455, Korea}

	\date{\today}
	
\begin{abstract}
Circadian rhythms in living organisms are temporal orders emerging from biochemical circuits driven out of equilibrium. 
Here, {\color{black}considering the KaiABC system, a minimal model in the synthetic biology, we study how the oscillation emerges from the circuit made of three Kai proteins and ATP alone}. 
The phase diagram constructed in terms of KaiC and KaiA concentrations reveals a narrowly bounded oscillatory phase{\color{black}, which naturally explains arrhythmia upon protein over-expression.}   
As dictated by the cost-precision trade-offs of the thermodynamic uncertainty relations, the presence of intrinsic noise, amplified in small systems, demands higher free energy cost to achieve greater rhythmic precision.
{\color{black}The cost-minimizing condition within the oscillatory phase is found to generate $\sim$21-hr rhythm, which is entrained to 24-hr environmental signals as long as the forcing amplitude is greater than $\sim 10$ \% of the metabolic rate.} 
An optimal level of intrinsic noise can {\color{black}also} induce oscillations {\color{black}even} beyond the Hopf bifurcation, effectively expanding the oscillatory phase. 
Our study clarifies how the physical factors, such as regulatory mechanism, energy cost, and stochastic noise contribute to the operation of biological clocks. 
\end{abstract}
	
\maketitle

\section{Introduction}
The  KaiABC clock in cyanobacteria, made of three core proteins, KaiA, KaiB, and KaiC, is arguably the simplest biochemical circuit for circadian rhythm that can be reconstituted \emph{in vitro}~\cite{nakajima2005Science,kageyama106MolCell}. 
Driven by $\sim$ 15 ATPs per day~\cite{kaicatp}, 
and 
regulated by KaiA and KaiB, KaiC protein
exhibits self-sustained $\sim$ 24-hr rhythms of change in its phosphorylation state even in the absence of {\color{black}an external} periodic driving
{\color{black}or transcription-translation feedback (TTF) regulation}
~\cite{kondo98science,nakajima2005Science}.

Experiments, interrogating the structures and interactions of Kai proteins and their mutants, have  {\color{black}offered hints on the molecular origin of the rhythmicity that is inherent in the KaiABC system without TTF regulation}~\cite{kondo98science,nakajima2005Science,michaelis_linearization}. 
Several theoretical studies, exploiting a set of coupled nonlinear ordinary differential equations (ODEs) with many state variables, have also addressed experimentally observed features, such as temperature compensation and synchronous oscillations in the ensemble via the KaiC-monomer exchange~\cite{emberly2006hourglass,vanWolde2007PNAS,hatakeyama2012generic,sasai22PLoSCB, Zhang2020NatPhys,fu2024temperature}. 
However, 
a large number of state variables in the existing ODE models and the complexity inherent to the nonlinearity make it highly demanding even to address the basic questions as to the design principles of a single KaiABC clock. 
For example, under what condition does the self-sustained oscillations emerge from the underlying circuit? 
How robust is the {\color{black}oscillatory rhythm} against the varying biochemical parameters? 

\begin{figure}
	\centering
	\includegraphics[width=0.72\linewidth]{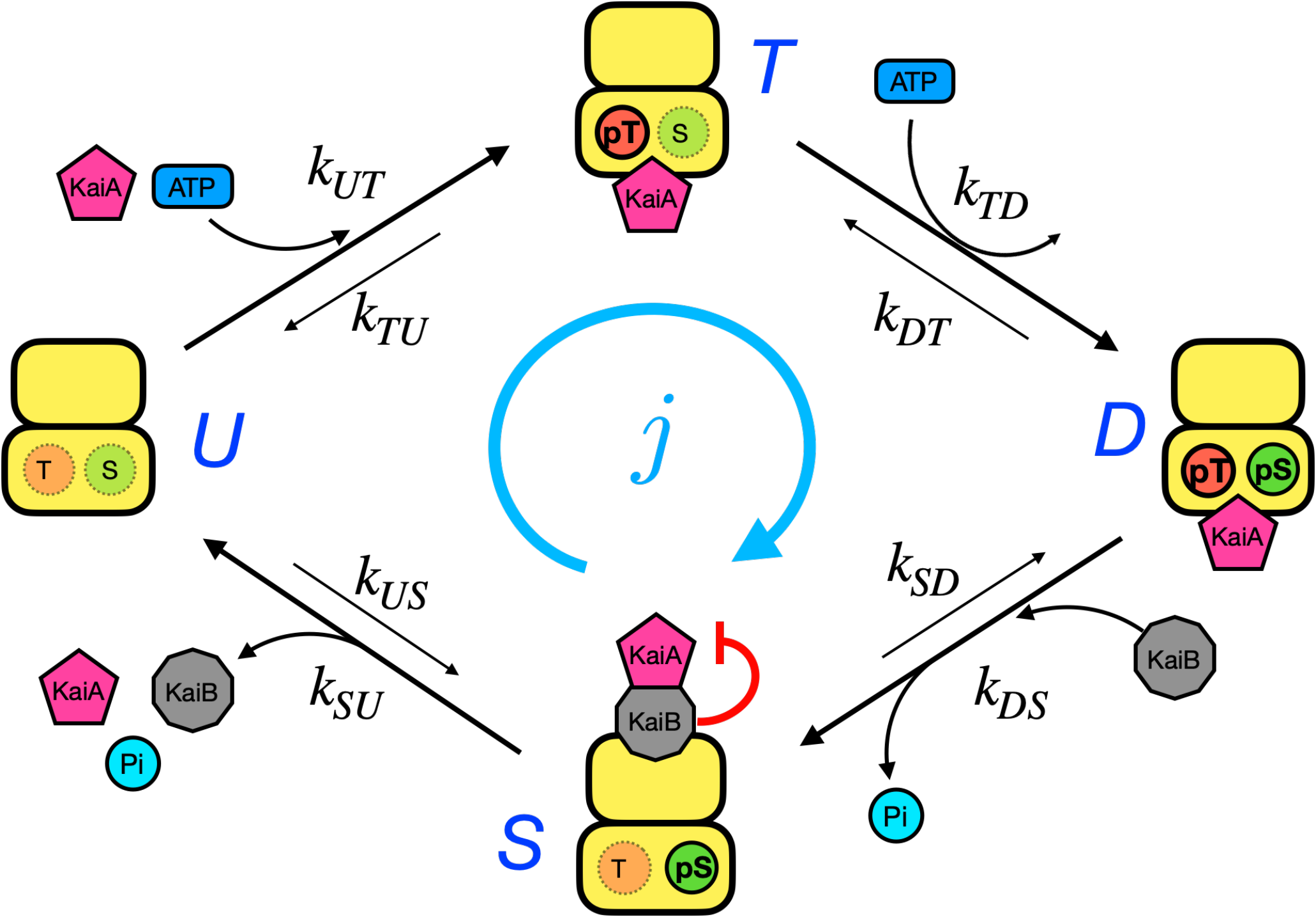}
	\caption{A kinetic scheme of the KaiC phosphorylation-dephosphorylation cycle for the ODE model studied here (see the main text for the details).  
	}
	\label{fig:reaction_diagram}
\end{figure}

{\color{black}Here, to put the KaiABC clock into the perspective of dynamical systems and to gain systems-level insights into various factors contributing to the generation of circadian rhythm, 
we employ a greatly simplified ODE model for the phosphorylation states of KaiC hexamer, developed by Rust \emph{et al.}~\cite{rust07Science}.} 
{\color{black}After describing the model, we perform the fixed point analysis of the ODE model for the KaiABC circuit to characterize its dynamical behavior in the deterministic limit. 
The resulting dynamical phase diagram of the circuit as a function of KaiA and KaiC concentrations reveals a narrowly bounded oscillatory phase.}
As dictated by the cost-precision trade-offs of the thermodynamic uncertainty relations (TURs)~\cite{barato2015PRL,marsland2019JRSI,Song2021JCP}, 
the free energy cost to generate a nonequilibrium process generally constrains the precision of the resulting dynamics subject to the noise~\cite{mehta2012PNAS,cao2015NatPhys,sartori2015free,Lan2012NaturePhysics}. {\color{black}Effect of intrinsic noise on the dynamics 
is unavoidable in cyanobacteria whose size is finite ranging from $\Omega\approx 10^0$ $\mu m^3$ to $\Omega\approx 10^6$ $\mu m^3$~\cite{Maranon24SciRep,Hahn12JBC}}. 
Quantifying such a relation {\color{black}and employing the uncertainty product factor from TUR as a measure of system size ($\Omega$)-independent free energy cost (dissipation) to generate oscillations amid fluctuations,} 
we identify the energetically most efficient condition for the KaiABC circuit to generate oscillations. 
Furthermore, {\color{black}we emphasize the effect of intrinsic noise on the aforementioned phase diagram calculated in the deterministic limit. Even} in the phase beyond the Hopf bifurcation lacking stable oscillations, the noise with a proper strength can help enhance a rhythmicity through the coherence resonance (CR)~\cite{gang1993PRL,pikovsky1997PRL,ushakov2005PRL}. 
By articulating these points for the KaiABC clock, 
{\color{black}we aim to offer} general insights into the design principles underlying biochemical oscillators.

\begin{figure}[t]
	\includegraphics[width=0.95\linewidth]{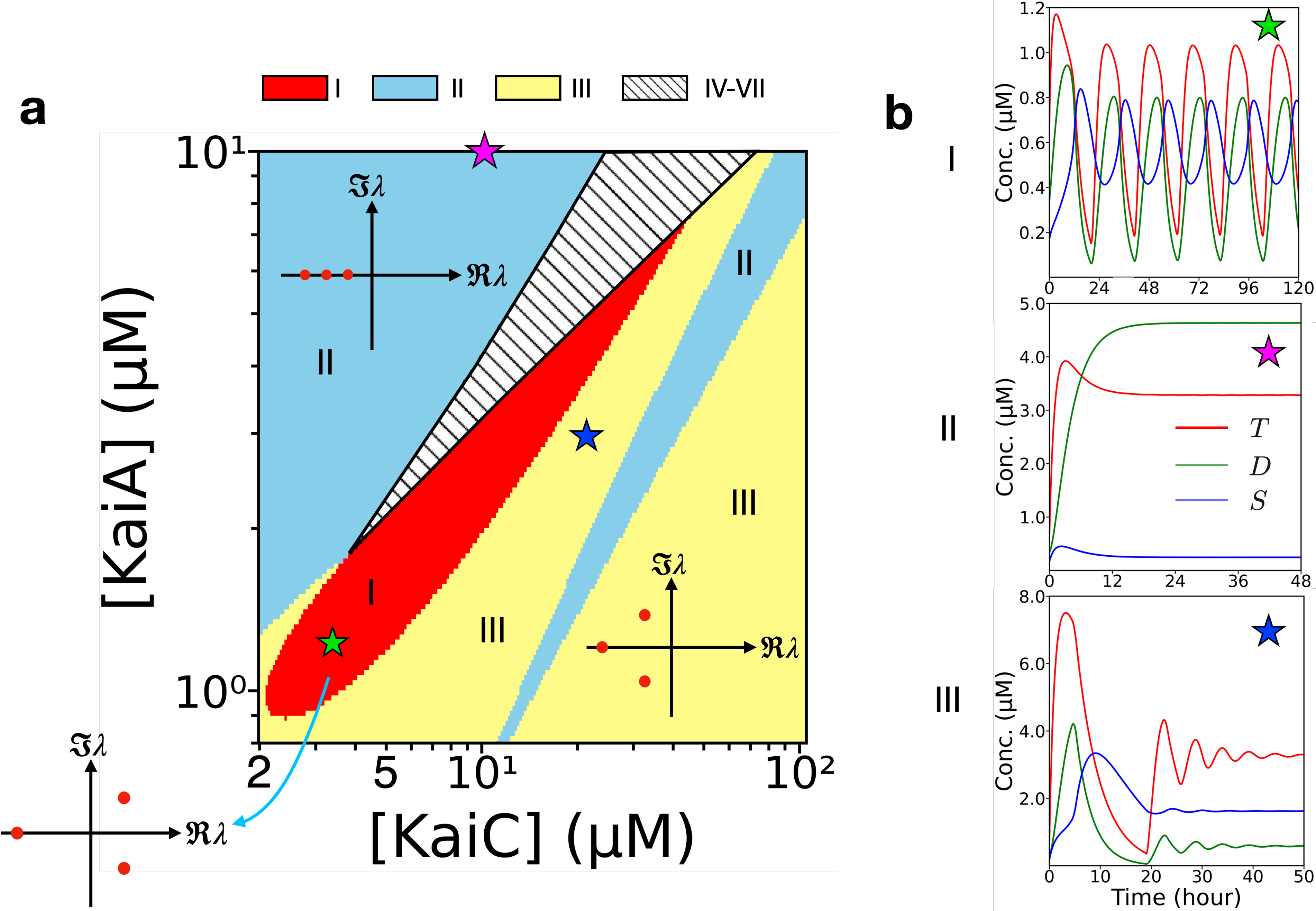}
	\caption{Dynamical phase diagram and trajectories from the  model of KaiABC clock.  
	(a) Phase I -- III classified based on the eigenvalue characteristics of the  fixed point. 
	More complicated structure of Phase IV -- VII is explained in the SM.  
	(b) Trajectories of $T$, $D$, and $S$ states of KaiC hexamer generated at $([{\rm KaiC}] (\mu{\rm M}),[{\rm KaiA}] (\mu{\rm M}))=(3.4,1.4)$ (green star) in Phase I, at $(10,10)$ (magenta star) in Phase II, and at $(20,3)$ (blue star) in Phase III. 
	} 
	\label{fig:phase_diagram}
\end{figure}

\section{Model}
KaiC is a hexameric protein, each monomer of which contains two phosphorylation sites (threonine 432 and serine 431)~\cite{nakajima2005Science}. 
Rust \emph{et al.}'s model~\cite{rust07Science} assumes an ordered KaiC phosphorylation cycle along with KaiA/KaiB-mediated effective rates, {\color{black}substantially} simplifying the complex allosteric dynamics of KaiC hexamer~\cite{iwasaki02PNAS,kitayama03EMBOJ,kageyama106MolCell,vanWolde2007PNAS,eguchi2008mechanism,Zhang2020NatPhys}. 
The model, formulated in terms of a set of nonlinear ODEs, is written as $\dot{\bf x}={\bf F}({\bf x})$ with {\color{black}only} three variables ${\bf x}=(T,D,S)$: 
\begin{align}
		\dot{T} &= k_{UT}(S)U + k_{DT}(S)D - \left[k_{TU}(S) + k_{TD}(S)\right]T \nonumber\\
		\dot{D} &= k_{TD}(S)T + k_{SD}(S)S - \left[k_{DT}(S) + k_{DS}(S)\right]D \nonumber\\
		\dot{S} &= k_{US}(S)U + k_{DS}(S)D - \left[k_{SU}(S) + k_{SD}(S)\right]S. 
	\label{eq:kaiabc_ode}
\end{align}
Eq.~\ref{eq:kaiabc_ode} represents the coherent evolution of a KaiC hexamer coordinated into four phosphorylation states as depicted in Fig.~\ref{fig:reaction_diagram}: (i) unphosphorylated state ($U$-KaiC or $U$), (ii) threonine-only-phosphorylated state ($T$-KaiC or $T$), (iii) serine-only-phosphorylated state ($S$-KaiC or $S$), and (iv) serine-threonine-phosphorylated state ($ST$-KaiC or $D$). 
Since the total concentration of KaiC is conserved without degradation {\color{black}within our relevant time of interest,} $U$ in Eq.~\eqref{eq:kaiabc_ode} is replaced with $U=[\mathrm{KaiC}] - T - D - S$. 
Catalyzed by KaiA, the cycle of KaiC (Fig.~\ref{fig:reaction_diagram}) begins with the phosphorylation of Thr432 (red circle, T$\rightarrow$pT), followed by  
the secondary phosphorylation at Ser431 (green circle, S$\rightarrow$ pS)~\cite{xu2004identification,egli2013loop}. 
Upon doubly phosphorylated, the KaiC undergoes a conformational change, and allows KaiB to bind and sequester KaiA, which induces the dephosphorylations of Thr432 and Ser431, and resets the cycle~\cite{rust07Science}.

In Eq.~\eqref{eq:kaiabc_ode}, 
the $S$-dependent transition rate from a state $X$ to $Y$ is modeled as 
\begin{align}
	k_{XY}(S) = k^{0}_{XY} + \frac{k^{A}_{XY}A(S)}{K_{1/2} + A(S)},  
    \label{eq:kXY}
\end{align}
where $A(S) = \text{max}\left(0, \left[\mathrm{KaiA}\right]-2S\right)$ corresponds to the concentration of free KaiA with 
the numerical factor 2 reflecting the 2:1 stoichiometry of the interaction between KaiA dimer and KaiC hexamer~\cite{hayashi2004stoichiometric}.
$k_{XY}^{0}$ is the basal rate of transition in the absence of free KaiA, and $k_{XY}^{A}$ in the second term denotes the maximal effect of KaiA on the rate constant. $K_{1/2}$ is the binding affinity between KaiA and KaiC. 
The expression of $k_{XY}(S)$ with $k_{UT}^A$, $k_{TD}^A >0$ models the KaiA-mediated positive regulation for phosphorylations, 
whereas $k_{XY}(S)$ with $k_{DS}^A$, $k_{SU}^A <0$   
effectively models the KaiB-mediated KaiA sequestration~\cite{brettschneider2010sequestration} which suppresses the KaiC autophosphorylation and activates the dephosphorylation process. 
Thus, the KaiB-regulation, 
accompanied with the accumulation of $S$-KaiC, is implicitly included in the model. 
The feedback mechanism arising from KaiA sequestration~\cite{brettschneider2010sequestration} is incorporated into $k_{XY}(S)$ in Eq.~\eqref{eq:kXY}, such that 
it enhances synchronization when a phase delay is present between oscillators
(see Fig.~\ref{fig:mixing} for the synchronization dynamics of two out-of-phase KaiABC oscillators upon mixing).

The explicit values of all the rate constants involving Eq.~\eqref{eq:kaiabc_ode} such as $k_{XY}^{0}$ and  $k_{XY}^{A}$, determined as the best fit parameters against experimental measurements~\cite{rust07Science}, are given in Table~\ref{table}. 

\begin{figure*}[t!]
	\centering
	\includegraphics[width=0.95\textwidth]{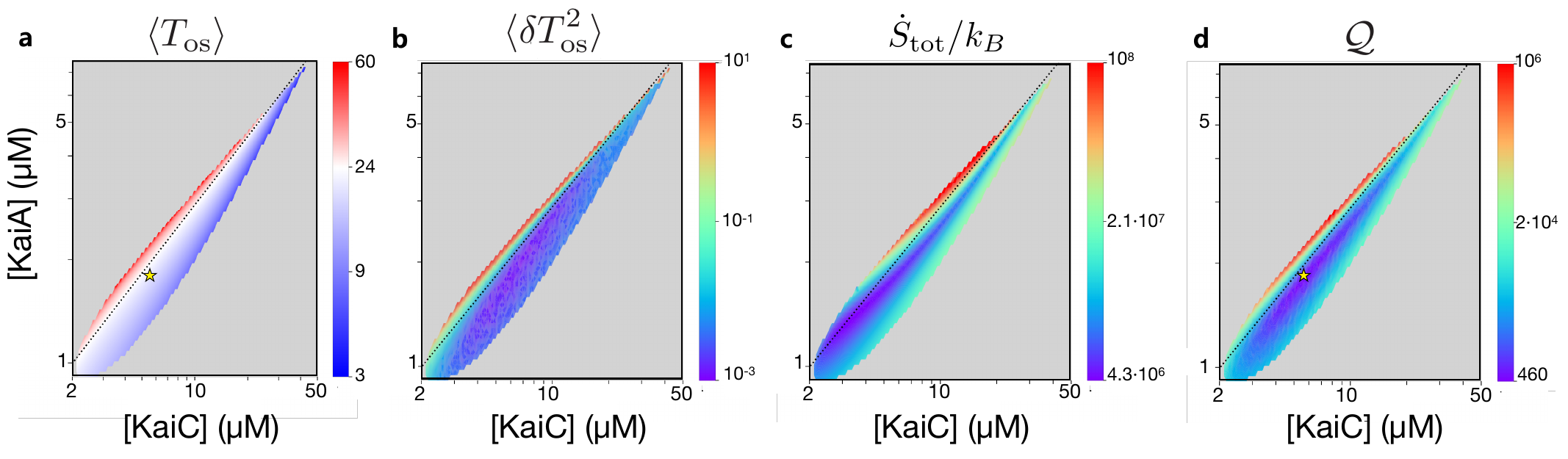}
	\caption{Dynamics in the oscillatory phase (Phase I) produced at {\color{black}a typical cell volume of cyanobacterium,} $\Omega = 1000~\mu\mathrm{m}^3$. 
	(a) Period of oscillation and (b) its variance, (c) entropy production, and (d) the uncertainty product. 
	{\color{black}Note that the magnitudes of $\langle\delta T^2_{\rm os}\rangle$ and $\dot{S}_{\rm tot}/k_B$ change with $\Omega$ as $\langle\delta T^2_{\rm os}\rangle\sim 1/\Omega$ and $\dot{S}_{\rm tot}/k_B\sim\Omega$ (see Fig.~\ref{fig:TUR_Omega}).}
	The yellow stars in Fig.~\ref{fig:TUR}a and \ref{fig:TUR}d mark the $\mathcal{Q}$-minimizing condition [KaiC]=5.71 $\mu$M and [KaiA]=1.87 $\mu$M, which leads to $\langle T_{\rm os}\rangle\simeq 21$ hr in (a).}
	\label{fig:TUR}
\end{figure*}

\section{Dynamical phase diagram}
The ODE model of KaiABC in Eq.~\eqref{eq:kaiabc_ode} represents  the time evolution of KaiC phosphorylation states at the deterministic limit, i.e., at $\Omega\rightarrow\infty$.  
To explore the full range of its dynamical behaviors, we vary [KaiC] and [KaiA], while assuming that the other parameters associated with the catalysis and protein-protein interactions, i.e., 
$k_{XY}^0$, $k_{XY}^A$, and $K_{1/2}$, are difficult to tune and held fixed.   
The linear stability analysis around the fixed points, ${\bf x}^\ast=(T^\ast,D^\ast,S^\ast)$ 
that satisfy ${\bf F}({\bf x}^\ast)=0$ 
in the range of $0~\leq~T^{*},~D^{*},~S^{*}~\leq~\mathrm{[KaiC]}$ {\color{black}(see Supplemental Materials)}, yields the dynamical phase diagram (Fig.~\ref{fig:phase_diagram}a). 

(i) Phase I is characterized by a single unstable fixed point whose Jacobian matrix has one negative real eigenvalue ($\lambda_1\in \mathbb{R}_{< 0}$) 
and a pair of complex conjugate eigenvalues with positive real part ($\lambda_{2,3}=\alpha\pm i\beta$ with $\alpha>0$). 
At [KaiC] = 3.4 $\mu$M and [KaiA] = 1.3 $\mu$M (green star in Fig.~\ref{fig:phase_diagram}a), {\color{black}which corresponds to the cost-minimizing point discussed later,} the trajectories of three phosphorylated states, namely, $T$, $D$, and $S$, exhibit stable periodic oscillations of $T_{\rm os}\sim$ 21-hr with phase lags (Fig~\ref{fig:phase_diagram}b-I).

(ii) Phase II and Phase III are characterized by a single stable fixed point, but each has different eigenvalue structure. 
For Phase II, $\lambda_{1,2,3}\in\mathbb{R}_{< 0}$, whereas for Phase III$, \lambda_1\in \mathbb{R}_{< 0}$ and $\lambda_{2,3}=\alpha\pm i\beta$ with $\alpha\in \mathbb{R}_{< 0}$.   
Trajectories generated in these phases (II and III) always converge to a stable fixed point; 
yet due to the difference in the eigenvalue structure, 
the convergence to the corresponding fixed point in Phase III is preceded by damped oscillations (blue star, Fig.~\ref{fig:phase_diagram}b-III), whereas the trajectories in Phase II exhibit direct damping without oscillation (magenta star, Fig.~\ref{fig:phase_diagram}b-II). 

The remaining phases demarcated by the hashed lines are characterized with three fixed points of various types, giving rise to complicated dynamics that requires more careful analysis. However, due to the presence of, at least, a single stable fixed point, trajectories in this region are convergent in the presence of noise at $t\rightarrow \infty$ (see SM Text and Fig.~\ref{fig:full_phase_diagram}).

Notably, Phase I is the only oscillatory phase. 
A few points are noteworthy. 
(i) The narrowly bounded oscillatory phase over the range of [KaiA] and [KaiC] straightforwardly accounts for experimentally observed arrhythmia (loss of rhythmicity or damping) when Kai proteins are either overexpressed or deleted from the system~\cite{rust07Science, kondo98science,kawamoto2020damped,xu2000circadian,iwasaki02PNAS,qin2010intermolecular,nakajima2010vitro}.  
Such sensitivity points to the importance of tightly regulated expression levels of KaiABC operon for maintaining the rhythmicity. 
(ii) For fixed [KaiA], the amplitude of oscillations changes non-monotonically with [KaiC] or $\langle T_{\rm os}\rangle$ (see Fig.~\ref{fig:amp}a and \ref{fig:amp}b), getting greater when [KaiA] and [KaiC] are higher (Fig.~\ref{fig:amp}a). 
(iii) Consistent with  experiments~\cite{kitayama03EMBOJ,iwasaki02PNAS,chang2011flexibility,nakajima2005Science, miwa20PNAS,ouyang98PNAS,martin06EMBOJ}, mutations that affect either the autokinase activity of KaiC or the binding affinity between KaiA and KaiC can alter the period of oscillations. 
Such effects can be incorporated to the current model by tuning the phosphorylation rates ($k_{UT}$ and $k_{TD}$) or the binding constant ($K_{1/2}$) in Eq.~\eqref{eq:kaiabc_ode} and Eq.~\eqref{eq:kXY} (see Fig.~\ref{fig:mutant_trace} for trajectories with altered periodicity and amplitude, and SM Text and Fig.~\ref{fig:mutation} for the changes in the corresponding phase diagrams).

\section{Dissipation constrains the regularity of oscillatory dynamics}
While high-dimensional dynamical systems may display complex behavior, suitable regulatory mechanisms can confine their long-term dynamics to a 2D invariant manifold exhibiting a limit cycle~\cite{kuznetsov1998elements}. 
The mean period of oscillation ($\langle T_{\rm os}\rangle$) calculated in Phase I
indicates that 
the 24-hr cycle is formed along the narrow parameter space, indicating a tight regulation of the clock (see Fig.~\ref{fig:period_detail} for the detailed view of the heat map of $\langle T_{\rm os}\rangle$). 
From both the numerical analysis of Eq.~\eqref{eq:kaiabc_ode} and its reduction to the Stuart-Landau form~\cite{kuznetsov1998elements} (see SM Text), 
we find a relation [KaiA] $\propto$ $\text{[KaiC]}^{\alpha}$ 
with $\alpha\approx 2/3$ satisfying the locus of $\langle T_{\rm os}\rangle\simeq 24$ hr in the heat map 
over the range of  [KaiA] $\approx (1-3)$ $\mu$M and [KaiC] $\approx (2-10)$ $\mu$M (dotted line in Fig.~\ref{fig:TUR}a. 
See also Figs.~\ref{fig:period_detail} and \ref{fig:scaling}). 
This is, in effect, consistent with the 1/3 KaiA to KaiC ratio for \emph{in vitro} oscillations~\cite{nakajima2010vitro}. 

Noting that the KaiABC clock is a nonequilibrium device exhibiting self-sustained oscillations under a constant drive of ATP, 
we put its rhythmic operation into the perspective of thermodynamic uncertainty relation (TUR) and assess its thermodynamic optimality~\cite{dechant2018JSM,Hwang2018JPCL,horowitz2019NaturePhys}. 
{\color{black}In particular, we adapt the TUR for the first passage time processes~\cite{gingrich2017PRL} and employ the relation to assess the cost-precision trade-off and identify the cost-minimizing condition for the operation of stochastic  clock~\cite{marsland2019JRSI,kim2021JPCB}.}
The uncertainty product $\mathcal{Q}$ of TUR for a stochastic clock is written as~\cite{kim2021JPCB}
\begin{align}
	\mathcal{Q} = \frac{\Delta S_{cyc}}{k_{B}} \frac{\left\langle \delta T_{\rm os}^{2} \right\rangle}{\langle T_{\rm os} \rangle^{2}} = \frac{\dot{S}_{tot}}{k_{B}} \frac{\left\langle \delta T_{\rm os}^{2} \right\rangle}{\langle T_{\rm os} \rangle}, 
	\label{eqn:Q}
	\end{align}
where $\Delta S_{cyc}$ is the entropy production per clock cycle, the mean and variance of oscillatory period are given by  $\langle T_{\rm os} \rangle$ and $\langle \delta T_{\rm os}^2 \rangle$, respectively, and 
$\dot{S}_{tot}=\Delta S_{cyc}/\langle T_{\rm os}\rangle$ denotes the mean entropy production rate. 
Eq.~\eqref{eqn:Q} dictates that 
for a given $\mathcal{Q}$, 
the cost to operate the clock ($\Delta S_{cyc}$) is counter-balanced with the relative error in the period ($\langle\delta T_{\rm os}^2\rangle^{1/2}/\langle T_{\rm os}\rangle$). 
The value of dimensionless factor $\mathcal{Q}$, which varies with the clock ``design" (parameters, network topology, and regulatory mechanism of the biochemical circuit), is used to assess the efficiency of the clock by linking irreversibility to measurable fluctuations~\cite{dechant2018JSM,Hwang2018JPCL} or to the power efficiency ($\eta=\text{output power}/\text{input power}$) of stochastic engines or molecular motors~\cite{pietzonka2016JSM,Song2021JCP,Rodriguez-Franco2025QRB}. 
A stochastic clock with a smaller $\mathcal{Q}$ value can be considered better designed in terms of the precision and energy expenditure. 
For dynamical processes exhibiting stochastic oscillations under \emph{constant} driving, 
$\mathcal{Q}$ is bounded as $\mathcal{Q}\geq \mathcal{Q}_{\rm min}^{\rm univ}=2$~\cite{barato2016cost}. 
For a type of cyclic network (Fig.~\ref{fig:reaction_diagram}), 
the universal bound is achieved close to equilibrium~\cite{barato2015PRL,barato2016cost,koyuk2019PRL}, although the  KaiABC circuit would cease to oscillate in such a condition.  
{\color{black}Thus, our objective here is to identify the condition that minimizes $\mathcal{Q}$ within the oscillatory phase.}

Evaluation of $\mathcal{Q}$ in Eq.~\ref{eqn:Q} is straightforward. 
The network of KaiC cycle depicted in Fig.~\ref{fig:reaction_diagram} is none other than a reversible unicyclic network. 
The mean entropy production rate $\dot{S}_{tot}$ in Eq.~\eqref{eqn:Q} is given by   
\begin{align}
\dot{S}_{tot}/k_B = \left(j_{+} - j_{-}\right)\ln\left(\frac{j_{+}}{j_{-}}\right) \cdot \text{[KaiC]} \cdot \Omega,
\label{eqn:Stot}
\end{align}
where 
$j_{+} = {\overline{k}}_{UT}{\overline{k}}_{TD}{\overline{k}}_{DS}{\overline{k}}_{SU}/\Sigma$ and $j_{-} = {\overline{k}}_{US}{\overline{k}}_{SD}{\overline{k}}_{DT}{\overline{k}}_{TU}/\Sigma$ with $\Sigma  
= {\overline{k}}_{UT}{\overline{k}}_{TD}{\overline{k}}_{DS} 
+ {\overline{k}}_{TD}{\overline{k}}_{DS}{\overline{k}}_{SU} 
+ {\overline{k}}_{DS}{\overline{k}}_{SU}{\overline{k}}_{UT} 
+ {\overline{k}}_{SU}{\overline{k}}_{UT}{\overline{k}}_{TD} 
+ {\overline{k}}_{DT}{\overline{k}}_{SD}{\overline{k}}_{US} + {\overline{k}}_{SD}{\overline{k}}_{US}{\overline{k}}_{TU} 
+ {\overline{k}}_{US}{\overline{k}}_{TU}{\overline{k}}_{DT} + {\overline{k}}_{TU}{\overline{k}}_{DT}{\overline{k}}_{SD}$~\cite{QianBook}. 
{\color{black}Here, it should be noted that ${\overline{k}}_{XY}$ denotes the rate constant averaged over the cycle period at cyclic steady states, i.e., ${\overline{k}}_{XY}=(1/T)\int_0^Tk_{XY}(\tau)d\tau$.}
The rate constants in Eq.~\eqref{eq:kaiabc_ode} are the functions of [KaiA] and [KaiC], and so are both the steady-state current 
$j=\overline{k}_{XY}X^{ss}-\overline{k}_{YX}Y^{ss}=j_+-j_-$ (Fig.~\ref{fig:reaction_diagram}) and the entropy production (Eq.~\eqref{eqn:Stot}). 
Note that the entropy production is an extensive quantity that increases with the number of KaiC proteins in the system ($N_{\rm KaiC}=[{\rm KaiC}]\times\Omega$, Eq.~\eqref{eqn:Stot}), i.e., $\dot{S}_{tot}\propto \Omega$ which is compensated by $\left\langle \delta T_{\rm os}^{2} \right\rangle \propto \Omega^{-1}$ {\color{black}while constituting the uncertainty product}.
$\langle T_{\rm os} \rangle$, determined by the dynamics confined in a 2D invariant manifold around the unstable fixed point, is independent of $\Omega$ (see Fig.~\ref{fig:TUR_Omega}). 
Taken together, the uncertainty product $\mathcal{Q}$ defined in Eq.~\eqref{eqn:Q} 
is an \emph{$\Omega$-independent} measure. 

When [KaiA] and [KaiC] are varied while holding all other parameters involving  catalytic power and protein-protein interactions (Table~\ref{table}),  
{\color{black}the $\mathcal{Q}$-minimizing condition KaiABC clock with $\mathcal{Q}_{\text{min}} \simeq \text{460}$ is formed at [KaiC] = 5.71 $\mu$M and [KaiA] = 1.82 $\mu$M, yielding the oscillatory period of $\langle T_{\rm os} \rangle \simeq 21$ hr 
(Figs.~\ref{fig:TUR}a and ~\ref{fig:TUR}d).} 
Similarly to error-correcting biochemical processes with large $\mathcal{Q}$ values that have to expend large free energy~\cite{Song2021JCP}, 
the KaiABC clock, severely underperforming the universal TUR bound ($\mathcal{Q}_{\text{min}} \simeq \text{460}\gg \mathcal{Q}_{\rm min}^{\rm univ}=2$), might be considered inefficient.   
Nevertheless, 
{\color{black}it should be noted that in order to create a temporal order from a given biochemical circuit,  
a finite amount of free energy consumption is inevitable, which necessarily drives the system away from the detailed balance condition corresponding to $\mathcal{Q}=\mathcal{Q}_{\rm min}^{\rm univ}=2$.   
For the case of KaiABC system, the cost-minimizing point to create an oscillation is formed at $\mathcal{Q}=\mathcal{Q}_{\text{min}}\simeq 460$ within the oscillatory phase, when [KaiA] and [KaiC] are the only two variables}.

Figure~\ref{fig:mutation} demonstrates a case in which $\mathcal{Q}_{\rm min}$ is slightly reduced ($\mathcal{Q}_{\rm min}\simeq 315$) upon increasing the binding affinity ($K_{1/2}=0.43\rightarrow 0.35$ $\mu{\rm M}$, Fig.~\ref{fig:mutation}d). 
In other cases, either the area of oscillatory phase shrinks (Figs.~\ref{fig:mutation}a,~\ref{fig:mutation}e), or when the area increases $\mathcal{Q}_{\rm min}$ increases significantly (Figs.~\ref{fig:mutation}b,~\ref{fig:mutation}c). 

\begin{figure}[t]
	\includegraphics[width=0.92\linewidth]{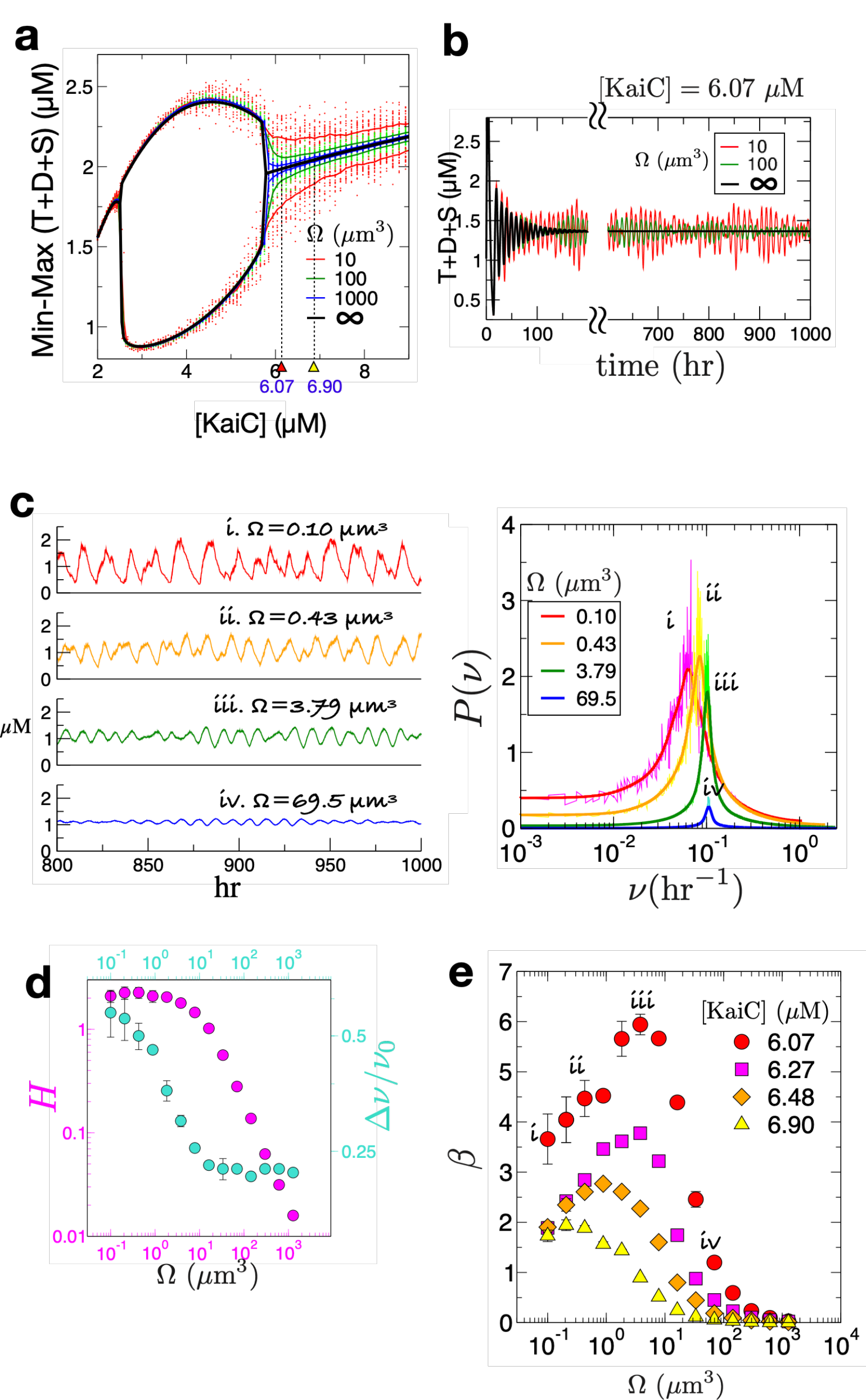}
	\caption{Noise-induced oscillation. 
	(a) Bifurcation diagram at $[{\rm KaiA}]=1.3$ $\mu$M. 
	Solid black line, blue, green, and red dots depict the minimum and the maximum values of $(T+D+S)$ for $\Omega\rightarrow\infty$, 1000, 100, and $10 \, \mu m^{3}$, respectively. 
	(b) Time evolutions of $(T+D+S)$ at $[{\rm KaiC}]=6.07$ $\mu$M. 
	(c) Noise-induced oscillations generated at [KaiC]$=6.07$ $\mu$M for varying $\Omega$s: $(i)$ $\Omega=0.10$, $(ii)$ $0.43$, $(iii)$ $3.79$, and $(iv)$ $69.5$ $\mu{\rm m}^3$ (left) and the corresponding power spectra $P(\nu)$s (right). 
	(d) Height ($H$) and width ($\Delta\nu/\nu_0$) of the resonant peak versus $\Omega$. (e) SNR ($\beta[=H/(\Delta \nu/\nu_0)]$) versus $\Omega$ for varying [KaiC]. 
	The values of $\beta$ calculated at different $\Omega$s in (c) are marked with $i$ to $iv$.} 
	\label{fig:bifurcation}
\end{figure}

\section{Noise-induced temporal order}
When [KaiC] is varied over $(2-10)~\mu$M at $\mathrm{[KaiA]=1.3~\mu M}$, passing through Phase I (Fig.~\ref{fig:phase_diagram}a),  
the stability of the fixed point undergoes a transition. 
Two supercritical Hopf bifurcation points, [KaiC]$_{cr}$ = 2.55 $\mu$M and 5.71 $\mu$M, are identified in the bifurcation diagram calculated at the deterministic limit, namely, at $\Omega\rightarrow\infty$  (Fig.~\ref{fig:bifurcation}a). 
Trajectories generated beyond the Hopf bifurcation point ($[{\rm KaiC}]=6.07$ $\mu$M), where deterministic oscillations are expected to vanish, 
display noisy oscillations at $\Omega = 10$ (red) and $100~\mu m^{3}$ (green) (Fig.~\ref{fig:bifurcation}b). 
The minimum and maximum values of the amplitude overlaid on the diagram,  
blur the sharp boundary obtained at $\Omega\rightarrow\infty$ (black line, Fig.~\ref{fig:bifurcation}a).

To study the effect of noise, whose intensity changes with the system size as $\propto 1/\sqrt{\Omega}$, more systematically, we examine the power spectra $P(\nu)[=\int C_{xx}(\tau)e^{-2\pi i \nu\tau}d\tau]$, the Fourier transform of autocorrelation function at steady state $C_{xx}(\tau)=\frac{1}{T}\int_0^Tx(t)x(t+\tau)dt$ with $x(t)=(T+D+S)(t)$, for varying $\Omega$ at [KaiC]$=6.07$ $\mu$M (Fig.~\ref{fig:bifurcation}b and ~\ref{fig:bifurcation}c).  
A resonant peak formed in $P(\nu)$ at $\nu=\nu_0 \sim 10^{-1}/$hr points to the {\color{black}$\sim10$ hr-}rhythmicity in the trajectories (Fig.~\ref{fig:bifurcation}c),  
{\color{black}although this deviates significantly from the $\sim$ 24 hr rhythm being discussed at the core of the oscillatory phase in the previous section.} 
The height ($H$) and width ($\Delta \nu/\nu_0$) of the resonant peak in $P(\nu)$ increase monotonically with the noise level (Fig.~\ref{fig:bifurcation}d), which are consistent with those {\color{black}previously} discovered near supercritical Hopf-bifurcation {\color{black}in Ref.~\cite{ushakov2005PRL}}. 
For [KaiC]$=6.07$ $\mu$M,  
the signal-to-noise ratio (SNR) of the resonant peak, i.e., the regularity of oscillations in time domain, quantified in terms of  
$\beta=H/(\Delta\nu/\nu_0)$ is maximized at an intermediate system size $\Omega_{\rm max}=\arg\max\limits_{\Omega}{\beta}\simeq 4$ $\mu$m$^3$ (Fig.~\ref{fig:bifurcation}e), 
which signifies that there is an optimal noise intensity (or $\Omega$ value) that yields the maximal regularity in oscillations (equivalently, the maximal SNR of the signal in frequency space).    
Small noise added to a stable trajectory is ineffective to induce oscillations, whereas large noise 
is also expected to hinder generation of oscillations with a regular period. 
When [KaiC] is varied away from the Hopf-bifurcation point ([KaiC]$_{cr}=5.71$ $\mu$M),  
the optimal noise level for the resonance shifts towards the smaller $\Omega$ (see the peak positions in Fig.~\ref{fig:bifurcation}e), suggesting that stronger noise is required to compensate for the increased distance from the bifurcation. 
Our finding that an optimal level of intrinsic noise can induce temporal order in a trajectory that is otherwise stable and non-oscillatory  corresponds to the \emph{coherence resonance} (CR) or \emph{stochastic resonance without periodic force}~\cite{gang1993PRL} that was also discovered for excitable systems~\cite{pikovsky1997PRL,li2007coupling}.

\begin{figure}[ht!]
		\includegraphics[width=1.0\linewidth]{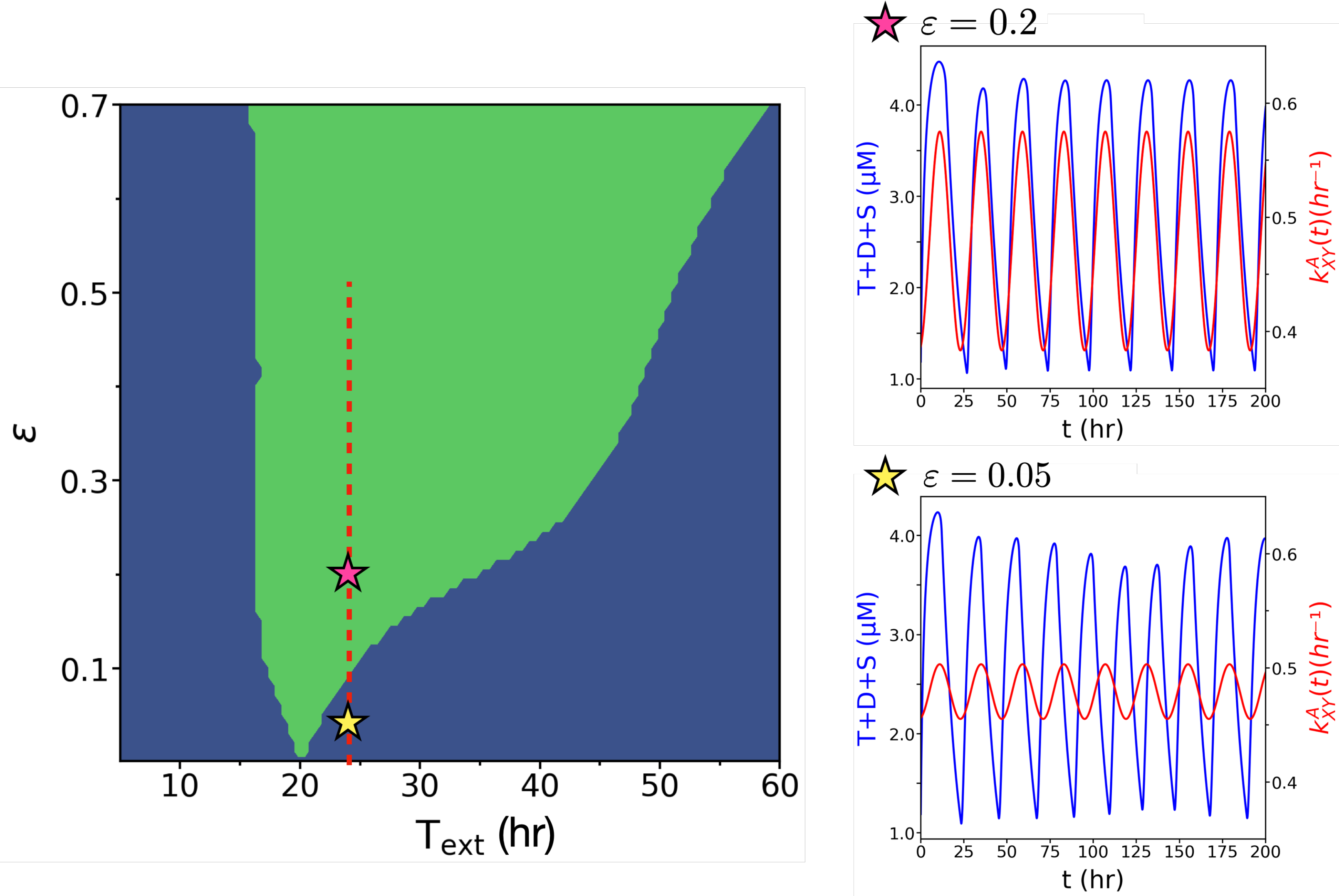}
		\caption{{\color{black}Entrainment of the $\sim$ 21-hr KaiABC rhythm at the cost-minimizing point to 24-hr environmental cues. 
		Based on Ref.~\cite{rust2011light} suggesting light-driven changes in metabolism, 
		we assume a modulation of KaiA-associated kinetic rate defined in Eq.~\eqref{eq:kXY}: $k_{XY}^A\rightarrow k_{XY}^A[1+\varepsilon\sin{(2\pi t/T_{\rm ext})}]$. 
		In the diagram of external forcing amplitude ($\varepsilon$) versus its period ($T_{\rm ext}$), 
		the entrainment of the intrinsic oscillation to the 24-hr cycle occurs in the regime depicted in green, corresponding to the \emph{Arnold tongue}. 
		An increasing $\varepsilon$ at $T_{\rm ext}=24$, depicted with the red dashed line, indicates that the entrainment occurs when $\varepsilon$ is greater than 10 \% of the intrinsic reaction rate ($k_{XY}^A$). 
		Shown on the right is the entrainment of $T+D+S$ dynamics (blue) under external periodic forcing (red) at $\varepsilon=0.2$ (top panel), contrasted with the absence of entrainment at $\varepsilon=0.05$ (bottom panel).}}
		\label{fig:entrainment}
		\end{figure}

\section{Discussion}
Consistent with the existing experimental reports~\cite{rust07Science, kondo98science,kawamoto2020damped,xu2000circadian,iwasaki02PNAS,qin2010intermolecular,nakajima2010vitro}, 
the phase diagram revealed in this study clarifies 
that the oscillatory phase of the KaiABC is tightly regulated over the relatively narrow range of biologically tunable parameters, [KaiA] and [KaiC]. 
This is contrasted with other biochemical circuits, such as the quorum sensing model of repressilator~\cite{garcia2004modeling}, that give rise to a broad oscillatory phase under double-negative feedback loops~\cite{kim2012mechanism}.  
{\color{black}For the KaiABC clock emerging from the circuit made of three Kai proteins without TTF regulation}, the parameter space for time-keeping is sensitively balanced by the competing actions between the positive and negative feedbacks of KaiA and KaiB. 

Within the oscillatory phase, the parameter space $\mathbf{c}=\left(\text{[KaiC], [KaiA]}\right)$ 
giving rise to $\langle T_{\rm os} \rangle \simeq 24$ hr (dotted line in Fig.~\ref{fig:TUR}a) and 
the condition $\mathbf{c}^{*} = \left( \text{[KaiC]}^\ast, \text{[KaiA]}^\ast \right)$ for 
the cost-minimizing period $\left\langle T_{\rm os} \right\rangle \left(\mathbf{c}^{*}\right) \simeq 21$ hr are distinct; yet, their separation in time is not significant. 
{\color{black}For an external forcing with 24-hr periodicity, 
it is expected that the 21-hr oscillation is entrained to the external cycle as long as the forcing amplitude is large enough~\cite{fang2023synchronization,pikovsky2002synchronization,heltberg2021tale,kim2024glycolytic}. 
The stable entrainment occurs, effectively putting the system into the ``Arnold tongue",  
when the forcing amplitude of sinusoidal perturbation ($\varepsilon$) added to the KaiA-associated metabolic rate is greater than 10 \% of metabolic rate, i.e., $\varepsilon\gtrsim 0.1$ 
(see Fig.~\ref{fig:entrainment} and its caption).} 

Our study on the minimal biochemical oscillator suggests that a large $\mathcal{Q}$-factor, signifying large free energy dissipation, is inevitable for generation of {\color{black}oscillatory dynamics} from biochemical circuits. 
{\color{black}Our finding that substantial amount of ATP per cycle has to be expended to create an oscillation even from a simplistic biochemical circuit, in fact, aligns well} with other biophysical machineries, such as molecular motors~\cite{Hwang2018JPCL,Song2021JCP,maggi2023thermodynamic,Rodriguez-Franco2025QRB}, biological error-correction~\cite{murugan2012speed,Song2020JPCL,mallory2020we}, molecular chaperones~\cite{jarmoskaite2021atp,song2022moderate}, and pattern formation during the early stage of embryogenesis~\cite{song2021cost,zhang2023free}. 

{\color{black}The results from this study on KaiABC clock are general, not necessarily specific to the Rust \emph{et al}'s model.  
The dynamic quantities required for calculating the uncertainty product ($\mathcal{Q}$), i.e., $\langle T_{\rm os}\rangle$, $\langle \delta T_{\rm os}^2\rangle$, and $\dot{S}_{tot}/k_B$, can be quantified entirely based on the experimentally measured time traces. 
For given dynamical trajectories, the period and its variance are straightforward to quantify, although extracting an accurate value of entropy production based on Eq.~\eqref{eqn:Stot} 
may involve a certain degree of subtlety, partly due to the level of coarse-graining that the Rust \emph{et al}'s model adopts. 
We assume that the trajectories generated under different conditions (e.g., at different values of [KaiC] and [KaiA]) are reasonable extrapolations of the model whose kinetic parameters are decided against the experimental measurement (Table~\ref{table}).}

The cost-optimal rhythmicity achieved despite substantial free-energy expenditure and the noise-induced temporal order revealed in this study suggest how
physics constrains the operation of stochastic clock in cells that emerges from biochemical circuit. 
Under the hood of dynamical systems~\cite{pikovsky2002synchronization} and stochastic thermodynamics~\cite{Seifert2012RPP}, 
these dynamical features can be regarded generic to biological clocks. 
They can, in principle, be uncovered in more complicated biochemical oscillators upon phase reduction and center manifold reduction of the associated dynamics~\cite{winfree2001book,kuramoto2003book}. 
Our study dissecting the capacity of biochemical circuits provides concrete physical insights into the principles governing self-sustained biological clocks and can potentially be extended to shed lights on the bioenergetics at the cellular scale~\cite{song2019CurrBiol,yang2021physical,di2024variance} and beyond~\cite{padamsey2023paying,jamadar2025metabolic}.  
Building on earlier efforts to construct synthetic gene oscillators with tunable or self-sustained dynamics~\cite{elowitz2000synthetic,stricker2008fast,potvin2016synchronous}, our findings can be of great use in guiding the rational design of synthetic oscillators.


\begin{acknowledgements}
We acknowledge the support from a KIAS individual grants CG097901 (YL) and CG035003 (CH) at the Korea Institute for Advanced Study. 
We thank the Center for Advanced Computation in KIAS for providing the computing resources.
\end{acknowledgements}

\bibliography{refs.bib,mybib1.bib}

\renewcommand{\thefigure}{S\arabic{figure}}
\renewcommand{\theequation}{S\arabic{equation}}
\renewcommand{\thetable}{S\arabic{table}}
\setcounter{figure}{0}
\setcounter{equation}{0}
\setcounter{section}{0}

\section*{Supplemental Material}
\section{Linear stability analysis} 
Dynamical systems $\dot{\mathbf{x}}= \mathbf{F}(\mathbf{x})$ in 3D, 
expanded around a fixed point $\mathbf{x}^{*} = \left(x^{*}, y^{*}, z^{*}\right)$ satisfying $F_{i}(x^{*}, y^{*}, z^{*}) = 0$ ($i = 1, 2, 3$),  yield a set of linear differential equations 
\begin{align}
	\delta{\dot{\textbf{x}}} = \mathcal{J}\left(\mathbf{x}^{*}\right) \cdot \delta{\textbf{x}},
	\label{eq:jacobian}
\end{align}
where $\mathcal{J}({\bf x}^\ast)$ is a Jacobian matrix evaluated at the fixed point.
As the amplitude of the fluctuation $\delta{\textbf{x}}$ varies over time 
as $\delta{\textbf{x}} \propto e^{\lambda t}$, the stability of the dynamical system is determined by the sign of the real part of eigenvalues, $\lambda$, obtained from the characteristic equation
\begin{align}
	\det\left(\lambda I - \mathcal{J}\left(\mathbf{x}^{*}\right)\right) = 0,
\end{align}
where $I$ is the $3\times3$ identity matrix, which yields 
\begin{align}
	a_{3}\lambda^{3} + a_{2} \lambda^{2} + a_{1} \lambda + a_{0} = 0,
	\label{eq:characteristic}
\end{align}
where $a_{3} = 1$, $a_{2} = -\Tr(\mathcal{J})$, $a_{1} = \Tr(\mathcal{J})^{2} - \Tr(\mathcal{J}^{2})$, $a_{0} = -\det\left(\mathcal{J}\right)$ with $\mathcal{J}\equiv \mathcal{J}({\bf x}^\ast)$. 

If the coefficients of Eq.~\ref{eq:characteristic} at $\mathbf{x}=\mathbf{x^{*}}$ satisfies the Routh-Hurwitz stability criterion  
\begin{align}
	\begin{split}
		&a_{i} > 0,\quad\quad  i=0, 1, 2, 3; \\
		&a_{2}a_{1} - a_{3}a_{0} > 0, 
	\end{split}
\end{align}
then all the real parts of eigenvalues are negative ($\real(\lambda_k)<0$ for $\forall k\in \{1,2,3\}$) and the fixed point $\mathbf{x^{*}}$ is stable.

	\section{Jacobian matrix}
	The local linear stability analysis of Eq.~\eqref{eq:kaiabc_ode} at the fixed point $\mathbf{x^{*}}=\left(T^{*}, D^{*}, S^{*}\right)$ gives rise to the Jacobian
	\begin{align}
			\mathcal{J}({\bf x}^\ast) = 
		\begin{pmatrix}
			\mathcal{J}_{11} & \mathcal{J}_{12} & \mathcal{J}_{13}\\
			\mathcal{J}_{21} & \mathcal{J}_{22} & \mathcal{J}_{23}\\
			\mathcal{J}_{31} & \mathcal{J}_{32} & \mathcal{J}_{33}
		\end{pmatrix}
\end{align}
		whose matrix elements are given as  
		\begin{widetext}
		\begin{align}
				\mathcal{J}_{11} & = -\left\{k_{UT}(S^\ast) + k_{TU}(S^\ast) + k_{TD}(S^\ast)\right\}  \nonumber\\
				\mathcal{J}_{12} & = k_{DT}(S^\ast) - k_{UT}(S^\ast)  \nonumber\\
				\mathcal{J}_{13} & = - \frac{ 2K_{1/2} \Theta([\mathrm{KaiA}] - 2S^\ast)}{ (K_{1/2} + A(S^\ast))^{2} } \left\{ k_{UT}^{A} [\mathrm{KaiC}] - (k_{UT}^{A} + k_{TU}^{A} + k_{TD}^{A}) T^\ast + (k_{DT}^{A}-k_{UT}^{A})D^\ast - k_{UT}^{A} S^\ast  \right\} - k_{UT}(S^\ast)  \nonumber\\
				\mathcal{J}_{21} &= k_{TD}(S^\ast) \nonumber\\
				\mathcal{J}_{22} &= -\left\{k_{DT}(S^\ast) + k_{DS}(S^\ast) \right\}  \nonumber\\
				\mathcal{J}_{23} &= - \frac{ 2K_{1/2} \Theta([\mathrm{KaiA}] - 2S^\ast)}{ \left( K_{1/2} + A(S^\ast) \right)^{2} } \left\{ k_{TD}^{A} T^\ast - \left( k_{DT}^{A} + k_{DS}^{A} \right) D^\ast + k_{SD}^{A} S^\ast \right\} + k_{SD}(S^\ast)  \nonumber\\
				\mathcal{J}_{31} &= -k_{US}(S^\ast)  \nonumber\\
				\mathcal{J}_{32} &= k_{DS}(S^\ast) - k_{US}(S^\ast)  \nonumber\\
				\mathcal{J}_{33} &= - \frac{ 2K_{1/2} \Theta([\mathrm{KaiA}] - 2S^\ast)}{ \left( K_{1/2} + A(S^\ast) \right)^{2} } \left\{k_{US}^{A}[\mathrm{KaiC}] - k_{US}^{A} T^\ast + \left( k_{DS}^{A} - k_{US}^{A} \right)D^\ast - \left( k_{SU}^{A} + k_{SD}^{A} + k_{US}^{A} \right) S^\ast \right\} \nonumber\\
							&\quad\,- \left( k_{SU}(S^\ast) + k_{SD}(S^\ast) + k_{US}(S^\ast) \right)
			\end{align}

	\section{Stochastic simulations using Gillespie algorithm}
	To perform stochastic simulations of the KaiC phosphorylation dynamics, we reformulate Eq.~\eqref{eq:kaiabc_ode} by explicitly writing the concentration $X=T$, $D$, $S$ in terms of the system size $\Omega$, such that $X = N_{X}/\Omega$, where $N_{X}$ is the number of molecular species. Thus, the set of ODEs are recast as
\begin{align}
			\frac{dN_{T}}{dt} &= k_{UT}(N_{S})N_{U} + k_{DT}(N_{S})N_{D} - k_{TU}(N_{S})N_{T} - k_{TD}(N_{S})N_{T} \nonumber\\
			\frac{dN_{D}}{dt} &= k_{TD}(N_{S})N_{T} + k_{SD}(N_{S})N_{S} - k_{DT}(N_{S})N_{D} - k_{DS}(N_{S})N_{D} \nonumber\\
			\frac{dN_{S}}{dt} &= k_{US}(N_{S})N_{U} + k_{DS}(N_{S})N_{D} - k_{SU}(N_{S})N_{S} - k_{SD}(N_{S})N_{S} \nonumber\\
			&k_{XY}(N_{S}) = k^{0}_{XY} + \frac{k^{A}_{XY}A(N_{S})}{K_{1/2} + A(N_{S})} 
			\label{eq:descrete_ode}
		\end{align}
where $A(N_{S}) = \textrm{max}\left(0, N_{\textrm{KaiA}} - 2 N_{S}\right)/\Omega$.
	\end{widetext}
	
	To include stochasticity in the simulation, we incorporate Gillespie algorithm~\cite{gillespie1} by denoting each transition corresponding to the arrow depicted in Fig.~\ref{fig:reaction_diagram} as $R_{\alpha}$  
	\begin{align}
			R_{1} &= k_{UT}(N_{S}) N_{U},\quad
			R_{2} = k_{DT}(N_{S}) N_{D},\nonumber\\
			R_{3} &= k_{TU}(N_{S}) N_{T},\quad
			R_{4} = k_{TD}(N_{S}) N_{T}, \nonumber\\
			R_{5} &= k_{SD}(N_{S}) N_{S},\quad
			R_{6} = k_{DS}(N_{S}) N_{D},\nonumber\\
			R_{7} &= k_{US}(N_{U}) N_{U},\quad
			R_{8} = k_{SU}(N_{S}) N_{S}, 
\end{align}
and assume that one of the transitions occurs following the Poisson statistics. 
The reaction time $\tau$ for Poisson process is given by 
	\begin{equation}
		\tau = \frac{1}{R} \ln{\frac{1}{r_{1}}}
		\end{equation}
	where $R = \sum\limits_{\alpha=1}^{8} R_{\alpha}$ and $r_{1}\in (0,1)$ is a random number drawn from an uniform distribution.
	To decide which transition to occur, we compute
	\begin{align}
		\alpha^\ast = \arg\min\limits_{\alpha}{\sum_{k=1}^{\alpha} R_{k}} > r_{2} \cdot R
		\label{eqn:rand2}
		\end{align}
	where $r_{2}$ is an another random number drawn from the uniform distribution. For instance, if $\alpha^\ast=2$ is selected from Eq.~\eqref{eqn:rand2}, 
we consider that the $D\rightarrow T$ transition ($R_2$) occurs at time $t$ to $t + \tau$, and update the number of chemical species in the system as $N_{T} \leftarrow N_{T} + 1$ and $N_{D} \leftarrow N_{D} - 1$.
	Iterating this procedure for desired time duration produces stochastic trajectories. 

\begin{figure}[ht!]
	\centering
	\includegraphics[width=\linewidth]{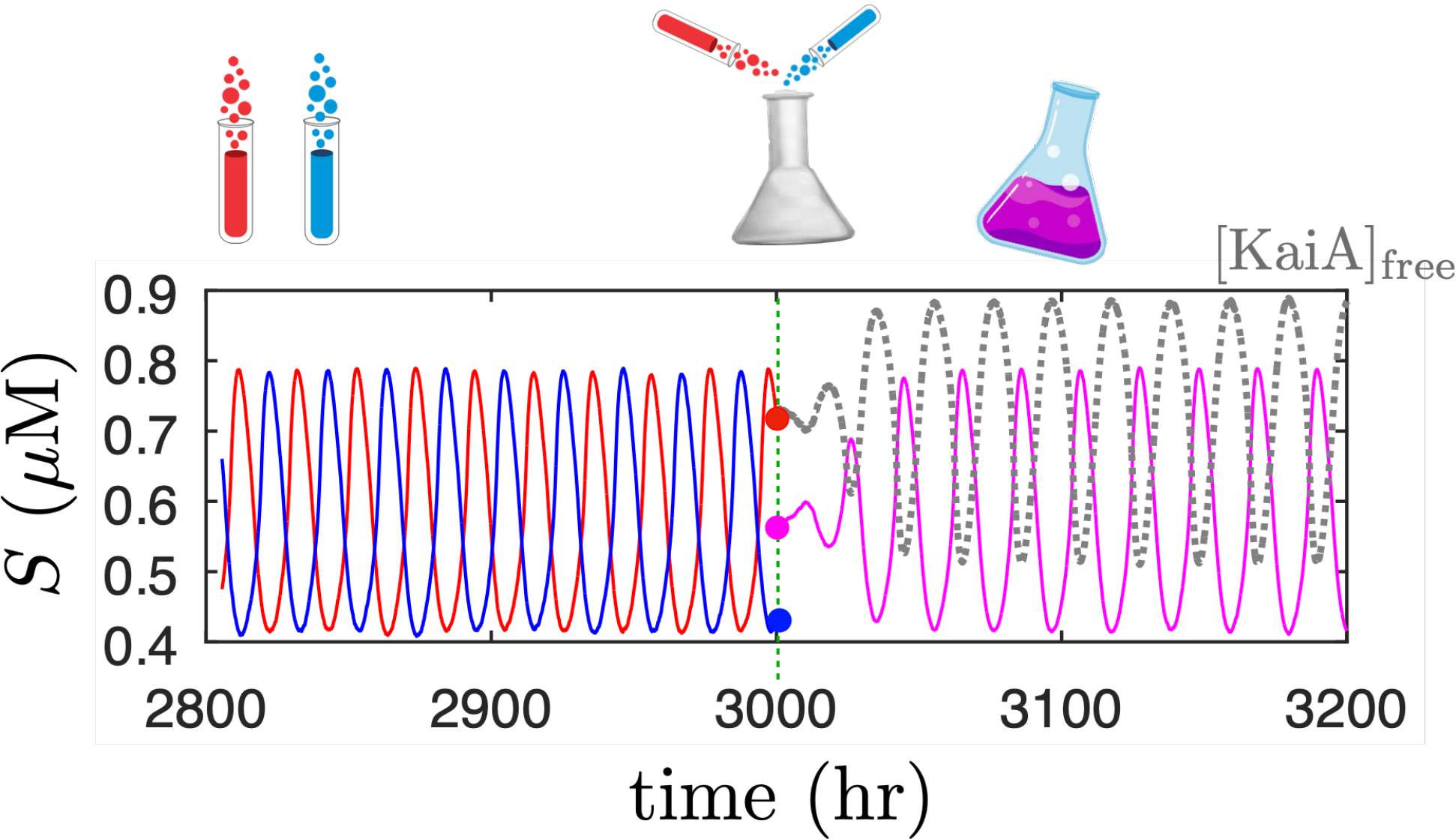}
	\caption{Restoration of the normal oscillatory dynamics of $S$ state (serine-only-phosphorylated state, $S$-KaiC) after a transient time of adjustment upon mixing two out-of-phase KaiABC oscillators.  The dotted line in grey depicts the concentration of free KaiA in the solution, [KaiA]$_{\rm free}=A(S)=\max{(0,[{\rm KaiA}]-S)}$.}
	\label{fig:mixing}
\end{figure}

\begin{figure*}[ht!]
\includegraphics[width=\linewidth]{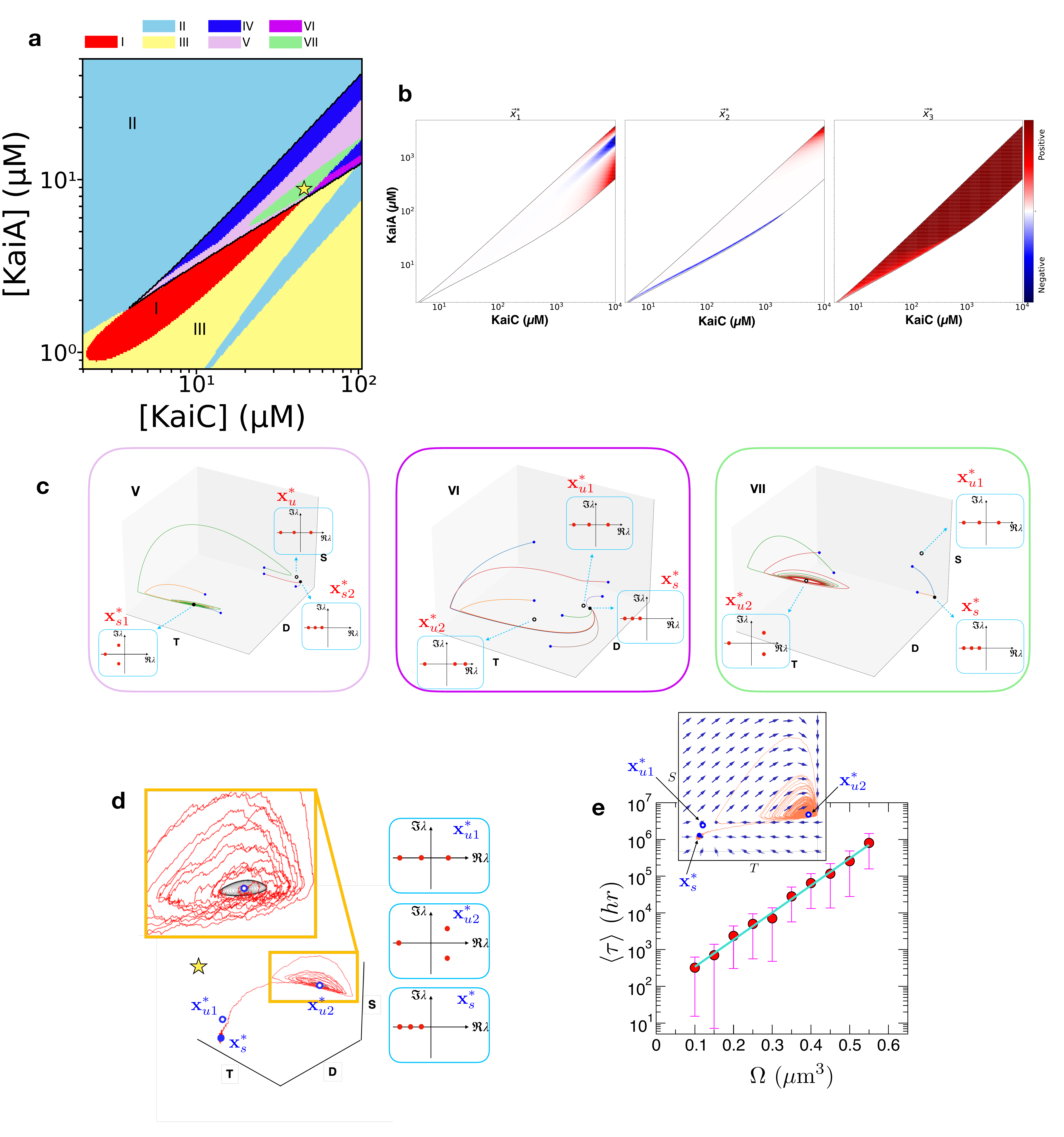}
\caption{
(a) Full phase diagram of KaiABC dynamics, where the detailed structure of the region unexplored in Fig.~\ref{fig:phase_diagram}a is shown explicitly and is demarcated using a  boundary with black line.   
(b) The value of $a_2a_1-a_3a_0$ for the three fixed points, $\vec{x}_1$, $\vec{x}_2$, and $\vec{x}_3$ in Phase IV, V, VI, and VII.
(c) The characteristics of the three fixed points in terms of the eigenvalues, and the corresponding trajectories generated in each phase.  
Stable and unstable fixed points are marked with black filled and empty circles in 3D space. 
The blue filled circles denote the initial starting point of each trajectory. 
Enclosed by the rectangles in cyan are the structures of three eigenvalues (red filled circles) depicted on the complex plane.
(d) Depicted are the deterministic (black, $\Omega\rightarrow \infty$) and stochastic trajectories generated at a finite volume (red, $\Omega=0.39$ $\mu$m$^3$) in 3D space of $(T,D,S)$ together with the structure of three eigenvalues (${\bf x}_{u1}^\ast$, ${\bf x}_{u2}^\ast$, ${\bf x}_{s}^\ast$) on the complex plane. 
(e) Exponentially growing mean first passage time from ${\bf x}_{u2}^\ast$ to ${\bf x}_{s}^\ast$, $\langle\tau\rangle\sim e^{\alpha\Omega}$ with $\alpha\approx 7.38$ (cyan line). 
	(Inset) The velocity field (blue arrows) projected on the $(T,S)$ plane and a trajectory generated from ${\bf x}_{u2}^\ast$ at $\Omega=0.6$ $\mu$m$^3$.  
}
\label{fig:full_phase_diagram}
\end{figure*}

\section{Synchronization of two KaiABC clocks upon mixing}
We consider two KaiABC systems under the same conditions [KaiC]$=3.4$ $\mu$M and [KaiA]$=1.3$ $\mu$M at $\Omega=1000$ $\mu$m$^3$, exhibiting $\sim$ 24 hr oscillation but with $\sim$ 12 hr phase shift. 
Upon mixing them at $t=3000$ hr, the combined system 
restores the 24 hr oscillation with the normal amplitude after a transient time of adjustment for $\approx 20$ hr (Fig.~\ref{fig:mixing}).

\section{Dynamics in the phases with three fixed points}
Phases IV, V, VI, and VII, the phase regions marked with hashed lines not discussed in the main text, are characterized by three fixed points. 
In Phase IV, all three fixed points are stable. 
Thus, any trajectory converges to one of the three fixed points at $t\gg 1$. 
On the other hand, at least one fixed point is unstable in  Phases V, VI, and VII.  
Thus, if there is a common parameter space where all the three fixed points are simultaneously unstable, chaotic dynamics can, in principle, arise.  
Nevertheless, Figure~\ref{fig:full_phase_diagram}b shows the value of $a_2a_1-a_3a_0$ of three fixed points $\vec{x}_1$, $\vec{x}_2$, and $\vec{x}_3$ over the varying [KaiA] and [KaiC] for Phase IV--VII, ruling out such a scenario of all three fixed points simultaneously violating the Routh-Hurwitz criterion. 
This means that at least one fixed point is always stable. 
Consequently, it is expected that trajectories, generated in a finite $\Omega$, i.e., in the presence of stochastic noise, converge to the stable fixed point. 

For Phases V and VI where the eigenvalues of unstable fixed points are real, trajectories always converge to a stable fixed point (Fig.~\ref{fig:full_phase_diagram}c). 
More complicated dynamics are observed in Phase VII where two unstable fixed points, ${\bf x}_{u1}^\ast$ and ${\bf x}_{u2}^\ast$, are characterized by a positive real eigenvalue and a pair of complex-conjugate eigenvalues with positive real part, respectively (see Fig.~\ref{fig:full_phase_diagram}c). 
Specifically, if the initial condition is chosen near ${\bf x}_{u1}^\ast$, 
the trajectories diverge from it and converge into the stable fixed point ${\bf x}_{s}^\ast$. 
In contrast, if trajectories originate from the point around ${\bf x}_{u2}^\ast$, 
they exhibit limit-cycle oscillations along the vortex field, $\dot{\bf x}=(\dot{T},\dot{D},\dot{S})$, surrounding ${\bf x}_{u2}^\ast$ in the absence of noise (or $\Omega\rightarrow\infty$: See the trajectory depicted by the black solid line in the inset that magnifies the trajectory in Fig.~\ref{fig:full_phase_diagram}d). 
In the presence of intrinsic noise at finite $\Omega$~\cite{gillespie1,vanKampen}), noisy limit-cycle oscillations generated around ${\bf x}_{u2}^\ast$ converge to ${\bf x}_{s}^\ast$ after a finite time, escaping from the vortex field (see the inset of Fig.~\ref{fig:phase_diagram}e).  
The mean first passage time from ${\bf x}_{u2}^\ast$ to ${\bf x}_{s}^\ast $ grows exponentially with $\Omega$ as $\langle\tau\rangle\sim e^{\alpha\Omega}$ (Fig.~\ref{fig:full_phase_diagram}e).

\section{Effects of mutation on the dynamics}
In comparison with the wild type, KaiC mutant with reduced phosphorylation rate requires higher concentration of KaiA to produce oscillatory dynamics, and shrinks the area of Phase I (Fig.~\ref{fig:mutation}a).  
Conversely, increasing the phosphorylation rates enlarges the oscillatory region and shifts it to a lower [KaiA] (Fig.~\ref{fig:mutation}b). 
The binding constant $K_{1/2}$ is another key factor that regulates the circadian period~\cite{iwasaki02PNAS}. 
Structural studies have shown that KaiA that binds the C-terminal tail of KaiC 
can interact with the ATP-binding cleft, and thus modulates the binding affinity~\cite{chang2011flexibility,dong16SR}. 
Moreover, ATP hydrolysis in KaiC promotes conformational changes that expose KaiA-binding sites, increasing the KaiA binding affinity~\cite{yunoki19LSA}. 
By varying the parameter $K_{1/2}$, we recalculate the phase diagram and find that increased binding affinity (decreased $K_{1/2}$) broadens the oscillatory region and enables limit-cycle behavior at lower KaiC concentrations (Fig.~\ref{fig:mutation}c). 
In contrast, an opposite effect is obtained for a reduced binding affinity (increased $K_{1/2}$), shifting the phase boundary towards higher KaiA and KaiC concentrations (Fig.~\ref{fig:mutation}d)~\cite{nishimura2002mutations}. 
Unless the perturbation to phosphorylation rate (or binding constant) is too severe, the oscillatory dynamics with $\sim$ 24 hr periodicity is preserved within the oscillatory phase. Fig.~\ref{fig:mutation}c demonstrates an extreme case where $k_{UT}$ and $k_{TD}$ are increased by 10 folds. In this case, the 24-hr periodicity is no longer retained within the oscillatory phase even though the area of the phase is enlarged, and the minimal value of $\mathcal{Q}$-factor is increased to $\mathcal{Q}_{\rm min}\sim 3\times 10^4$.

\section{Approximation of limit cycle dynamics in Phase I to the Stuart-Landau equation}
For each set of parameters ([KaiC],[KaiA]), the mean period of oscillations $\langle T_{\rm os}\rangle$ is calculated numerically by analyzing the resulting trajectory at the steady state (Fig.~3a). 

Since our limit cycle dynamics is realized by the trajectories that settle down in 2D invariant manifold, it is, in principle, possible to consider approximating Eq.~\eqref{eq:kaiabc_ode} to the Stuart-Landau equation of $z=Re^{i\phi}$ in the complex plane~\cite{kuznetsov1998elements}
\begin{equation}
    \frac{dz}{dt} = \left(\sigma_{r} + i\,\sigma_{i}\right)z + 
    \frac{l}{2}z|z|^{2}+\mathcal{O}(|z|^4),
\end{equation}
where $\sigma_{r}$ and $\sigma_{i}$ are the real and imaginary parts of a complex-valued eigenvalue of the Jacobian matrix $J$ at the fixed point (Eq.~\eqref{eq:jacobian}), acquired near the Hopf bifurcation point satisfying $\sigma_r\approx 0$. 
The complex Lyapunov coefficient, $l=l_r+il_i$, is calculated by 
\begin{align}
l&=\frac{1}{2}\langle \vec{p},\vec{C}(\vec{q},\vec{q},\vec{q}^\ast)\rangle
-\langle \vec{p},\vec{B}(\vec{q},J^{-1}\vec{B}(\vec{q},\vec{q}^\ast))\rangle\nonumber\\
&+\frac{1}{2}\langle\vec{p},(2i\omega_0I-J)^{-1}\vec{B}(\vec{q},\vec{q})\rangle
\end{align}
where $\langle\cdot,\cdot\rangle$ represents the inner product, $\vec{p}$ and $\vec{q}$ denote the left and right-eigenvectors for the above-selected eigenvalue satisfying 
$\langle\vec{p},\vec{q}\rangle=1$, $\vec{q}^\ast$ is the complex conjugate of $\vec{q}$, and $\omega_0$ is the value of $\sigma_i$ in the limit of $\sigma_r\rightarrow 0$. 
Each component of $\vec{B}=(B_1,B_2,B_3)$ and $\vec{C}=(C_1,C_2,C_3)$ 
is computed as 
\begin{align}
B_i(u,v)&=\sum_{j,k}\frac{\partial^2 F_i({\bf x})}{\partial x_j\partial x_k}\Big|_{{\bf x}={\bf x}^\ast}u_ju_k,\nonumber\\
C_i(u,v,w)&=\sum_{j,k,l}\frac{\partial^3 F_i({\bf x})}{\partial x_j\partial x_k\partial x_l}\Big|_{{\bf x}={\bf x}^\ast}u_ju_ku_l. 
\end{align}

Then, the amplitude ($R$) and the phase ($\phi$) evolve in time, obeying 
$\dot{R}=\sigma_rR-(l_r/2)R^3$ and $\dot{\phi}=\sigma_i-(l_i/2)R^2$, respectively. 
In this case, the amplitude $R(t)$ changes with time as 
\begin{align}
R(t)=\left[\frac{l_r}{2\sigma_r}+\left(|R(0)|^{-2}-\frac{l_r}{2\sigma_r}\right)e^{-2\sigma_rt}\right]^{-1/2}
\end{align}
and settles to $R(\infty)\rightarrow (2\sigma_r/l_r)^{1/2}$. 
On the other hand, the phase variable $\phi(t)$ increases with time $t$ as 
\begin{align}
\phi(t)&=\phi(0)+\sigma_it\nonumber\\
&-\frac{l_i}{2l_r}\ln{\left[1+\frac{|R(0)|^2l_r}{2\sigma_r}(e^{2\sigma_rt}-1)\right]}
\end{align}
and this expression offers the period of oscillation at steady state ($t\gg 1/2\sigma_r$), 
\begin{align}
\langle T_{\rm os}\rangle= \frac{2\pi}{\sigma_i-(l_i/l_r)\sigma_r}
\end{align}
Figure~\ref{fig:scaling} shows two panels:  
the panel on the left is from the numerical solution of ODEs, and the panel on the right is obtained by casting the original ODEs into the Stuart-Landau equation. 

Note that all the quantities discussed above for the Stuart-Landau equation, i.e., $\sigma_r$, $\sigma_i$, $l_r$, $l_i$, and so forth, are the values that can, in principle, be expressed symbolically in terms of [KaiC] and [KaiA]. 
The expressions, however, are algebraically too cumbersome to handle, and not particularly illuminating. 
Thus, to obtain the heatmap from the Stuart-Landau equation, 
we had to rely on numerics as well. 

Overall, the two heatmaps are found in reasonable agreement over the range of 
${\rm [KaiC]} = (2-5)$ $\mu{\rm M}$ and 
${\rm [KaiA]} = (1-2)$ $\mu{\rm M}$. 
Furthermore, the data points of ([KaiC],[KaiA]) that satisfy $|\langle T_{\rm os}\rangle({\rm [KaiC]},{\rm [KaiA]})-24\text{ }{\rm hr}|<0.5\text{ }{\rm hr}$ 
nicely overlap on the scaling relation of 
${\rm [KaiA]} \propto {\rm [KaiC]}^{2/3}$ (dotted line in Fig.~\ref{fig:scaling}), which lends support to the approximation of Eq.~\eqref{eq:kaiabc_ode} to the Stuart-Landau form.

\begin{figure*}[htbp]
	\includegraphics[width=1.0\linewidth]{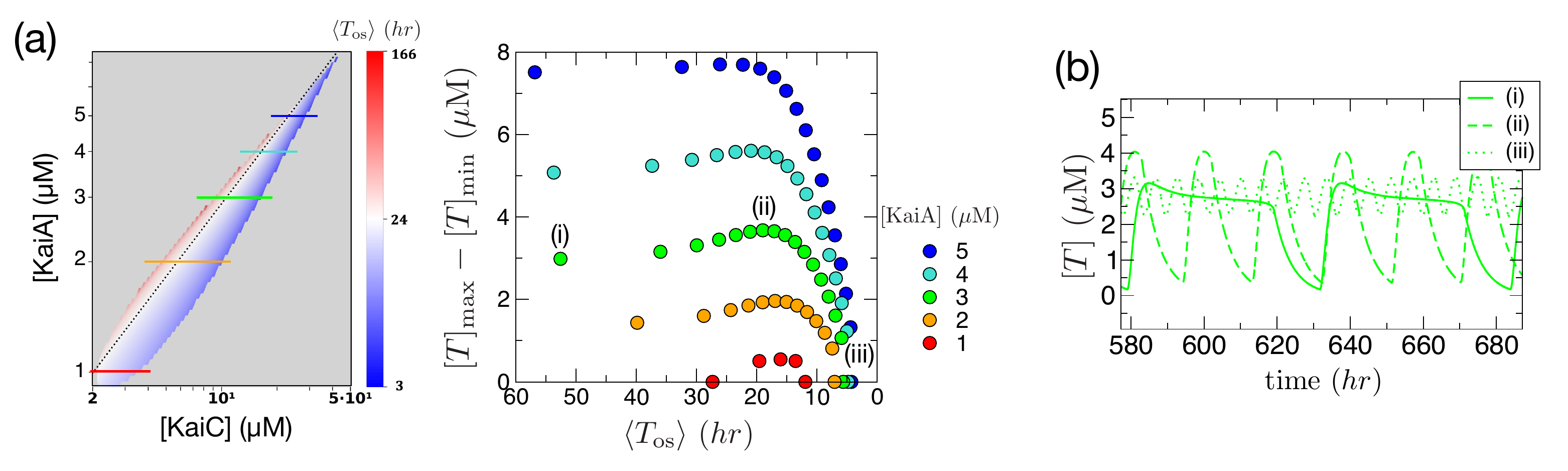}
	\caption{(a) Oscillation amplitude of $T$ state ($[T]_{\rm max}-[T]_{\rm min}$) as a function of mean period of oscillations (right) when Phase I is sliced at [KaiA]$=1,2,\ldots, 5$ $\mu$M (left). (b) Trajectories of $T$ state exhibiting oscillations with different amplitudes at (i) [KaiC]=9.27 $\mu$M, (ii) [KaiC]=12.2 $\mu$M, and (iii) [KaiC]=16.6 $\mu$M that are marked on the plot in (a).}
		\label{fig:amp}
	\end{figure*}

\begin{figure*}[htbp]
		\includegraphics[width=\linewidth]{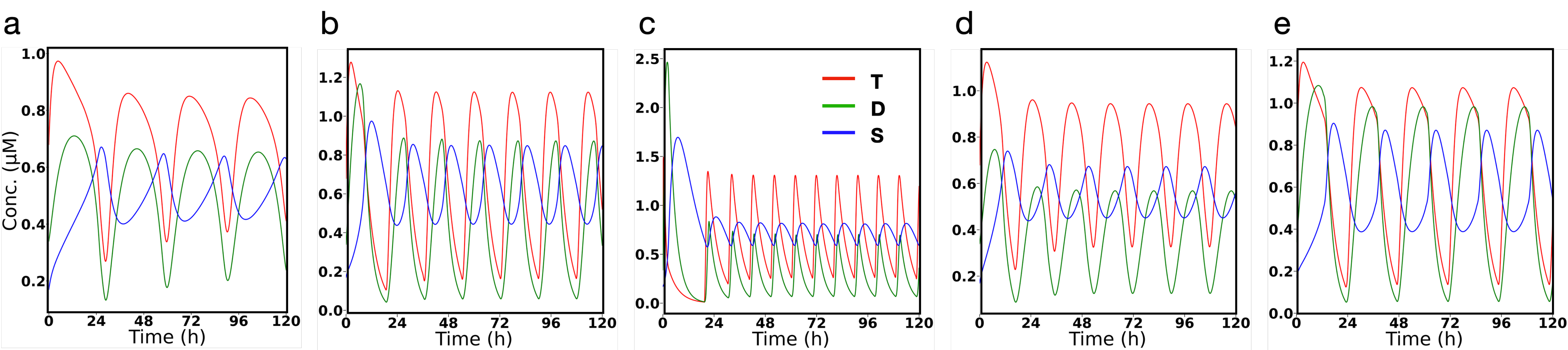}
		\caption{The oscillatory dynamics produced at [KaiC]=3.4 $\mu$M and [KaiA]=1.3 $\mu$M modulated by the changes made in the parameters ($k_{UT}$, $k_{TD}$, and $K_{1/2}$). 
		(a) Elongated oscillation period (30 hours) as a result of the changes, $k_{UT}\rightarrow 0.56\times k_{UT}$ and $k_{TD}\rightarrow 0.56\times k_{TD}$. 
		(b) Shortened period (17 hours) as a result of the changes, $k_{UT}\rightarrow 1.5\times k_{UT}$ and $k_{TD}\rightarrow 1.5\times k_{TD}$. 
		(c) Shortened period (10 hours) as a result of the changes, $k_{UT}\rightarrow 10\times k_{UT}$ and $k_{TD}\rightarrow 10\times k_{TD}$. 
		(d) Elongated period (24 hours) as a result of the change, $K_{1/2}=0.43$ $\mu$M $\rightarrow 0.35$ $\mu$M. 
		(e) Shortened period (18 hours) as a result of the change, $K_{1/2}=0.43$ $\mu$M $\rightarrow 0.60$ $\mu$M. 
		The period of oscillation for the unperturbed system   
	is $\sim$ 21 hours.}
	\label{fig:mutant_trace}
		\end{figure*}

	\begin{figure*}[t]
	\includegraphics[width=\linewidth]{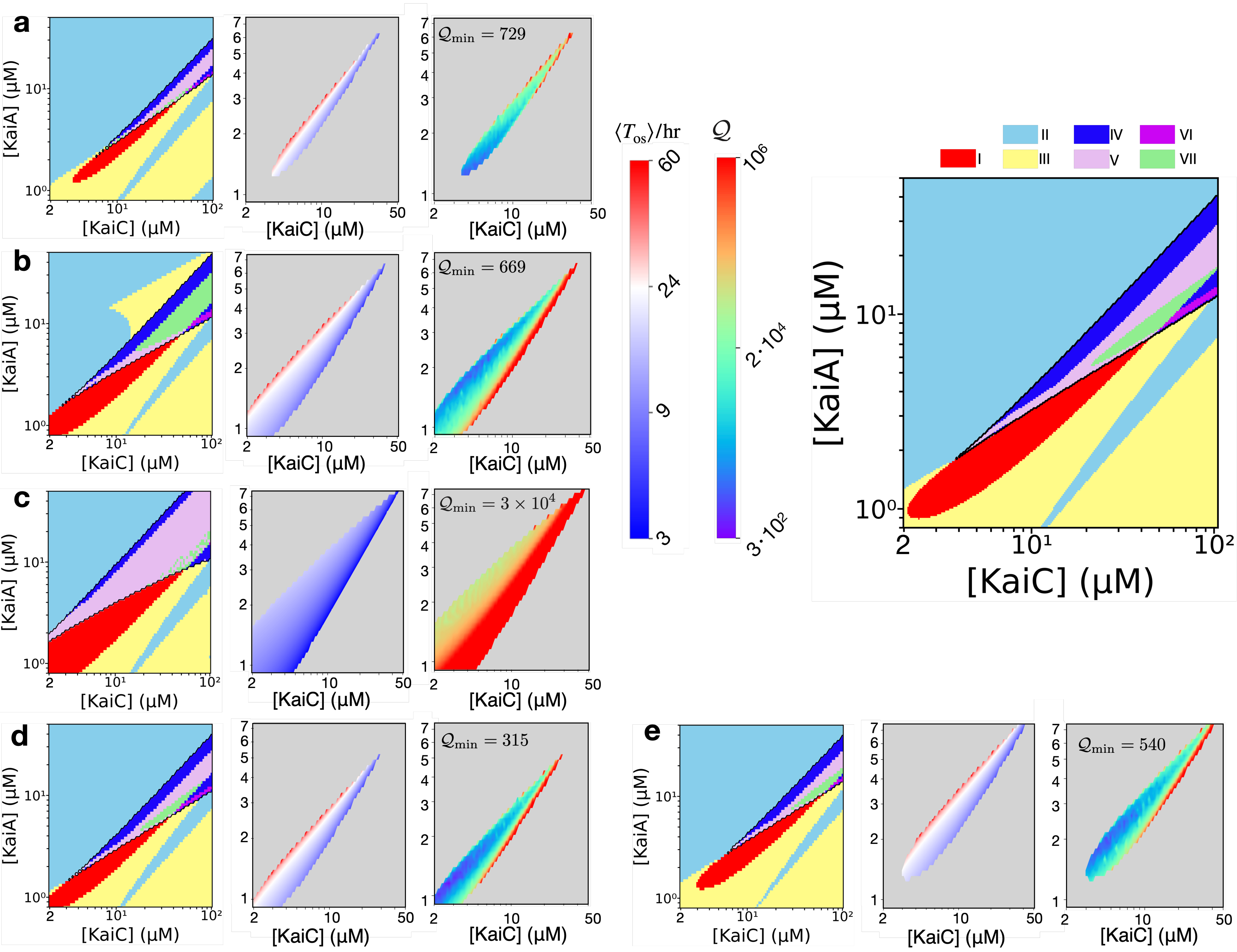}
	\caption{Effect of mutations on the original phase diagram (the main panel on the right) as a result of the modification of the rate or binding constants:
	(a) $k_{UT}\rightarrow 0.56\times k_{UT}$ and $k_{TD}\rightarrow 0.56\times k_{TD}$, 
	(b) $k_{UT}\rightarrow 1.5\times k_{UT}$ and $k_{TD}\rightarrow 1.5\times k_{TD}$, 
	(c) $k_{UT}\rightarrow 10\times k_{UT}$ and $k_{TD}\rightarrow 10\times k_{TD}$, 
	(d) $K_{1/2}=0.43$ $\mu$M $\rightarrow 0.35$ $\mu$M, 
	(e) $K_{1/2}=0.43$ $\mu$M $\rightarrow 0.60$ $\mu$M (see Fig.~\ref{fig:mutant_trace} for the  trajectories). 
	Each subfigure shows the phase diagram modified under  the mutation (left), the resulting heat map of $\langle T_{\rm os}\rangle([{\rm KaiC}],[{\rm KaiA}])$ in Phase I (middle), and the heat map of $\mathcal{Q}$ together with the value of $\mathcal{Q}_{\rm min}$ (right).  }
	\label{fig:mutation}
\end{figure*}	
			
\begin{figure*}
    \centering
    \includegraphics[width=0.5\linewidth]{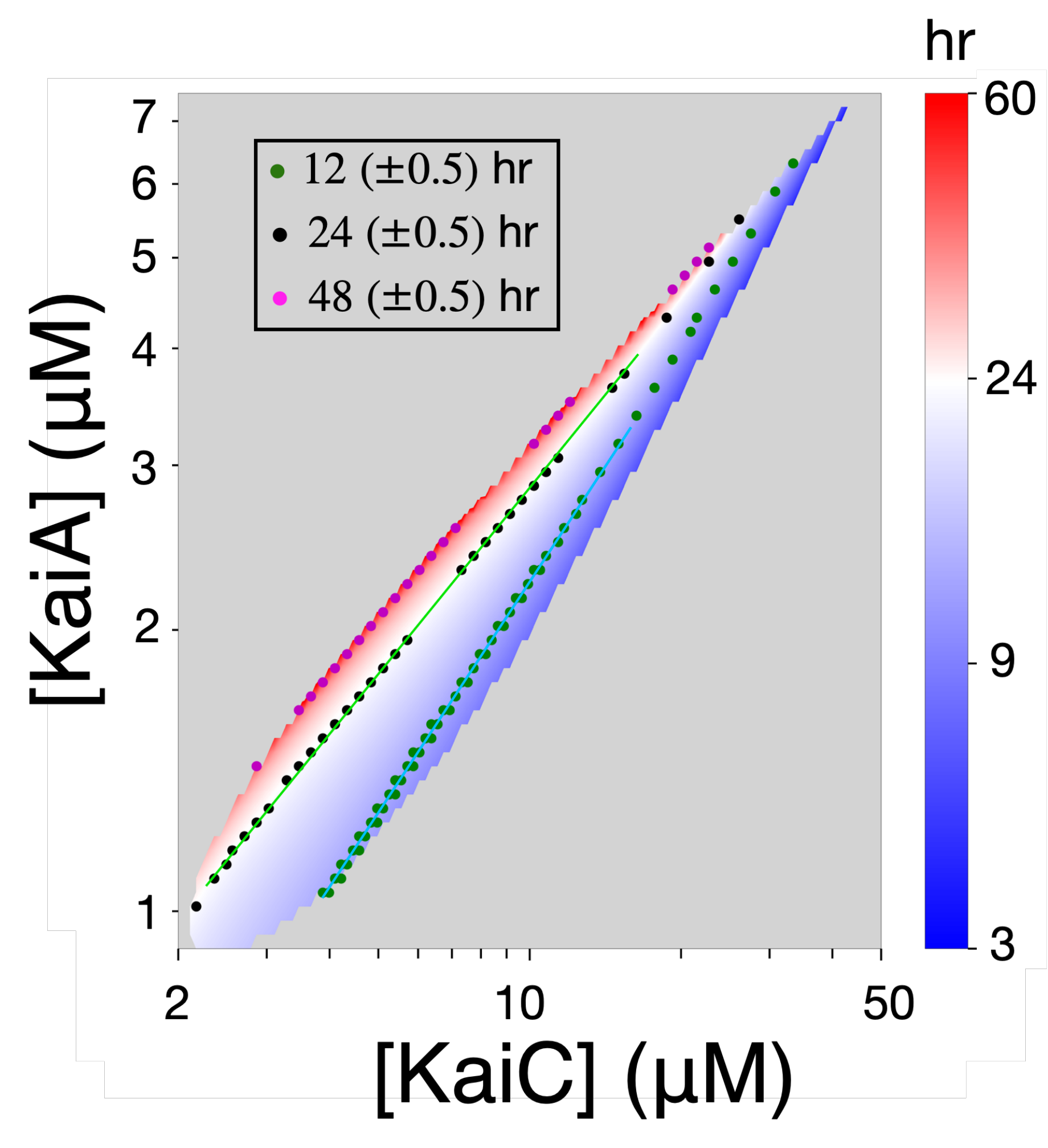}
    \caption{Detailed view of the heat map of mean oscillation period $\langle T_{\rm os}\rangle([{\rm KaiA}],[{\rm KaiC}])$.  
    Marked on the heat map are a set of concentrations $([{\rm KaiA}],[{\rm KaiC}])$ satisfying $\langle T_{\rm os}\rangle=12$, 24, and 48 hr. 
    The green and cyan lines correspond to the relations of 
    $[{\rm KaiA}]\propto[{\rm KaiC}]^{\alpha}$ with $\alpha\simeq 2/3$ and $4/5$ for $\langle T_{\rm os}\rangle=24$ and 12 hr respectively. 
    }
    \label{fig:period_detail}
\end{figure*}

\begin{figure*}
    \centering
    \includegraphics[width=0.7\linewidth]{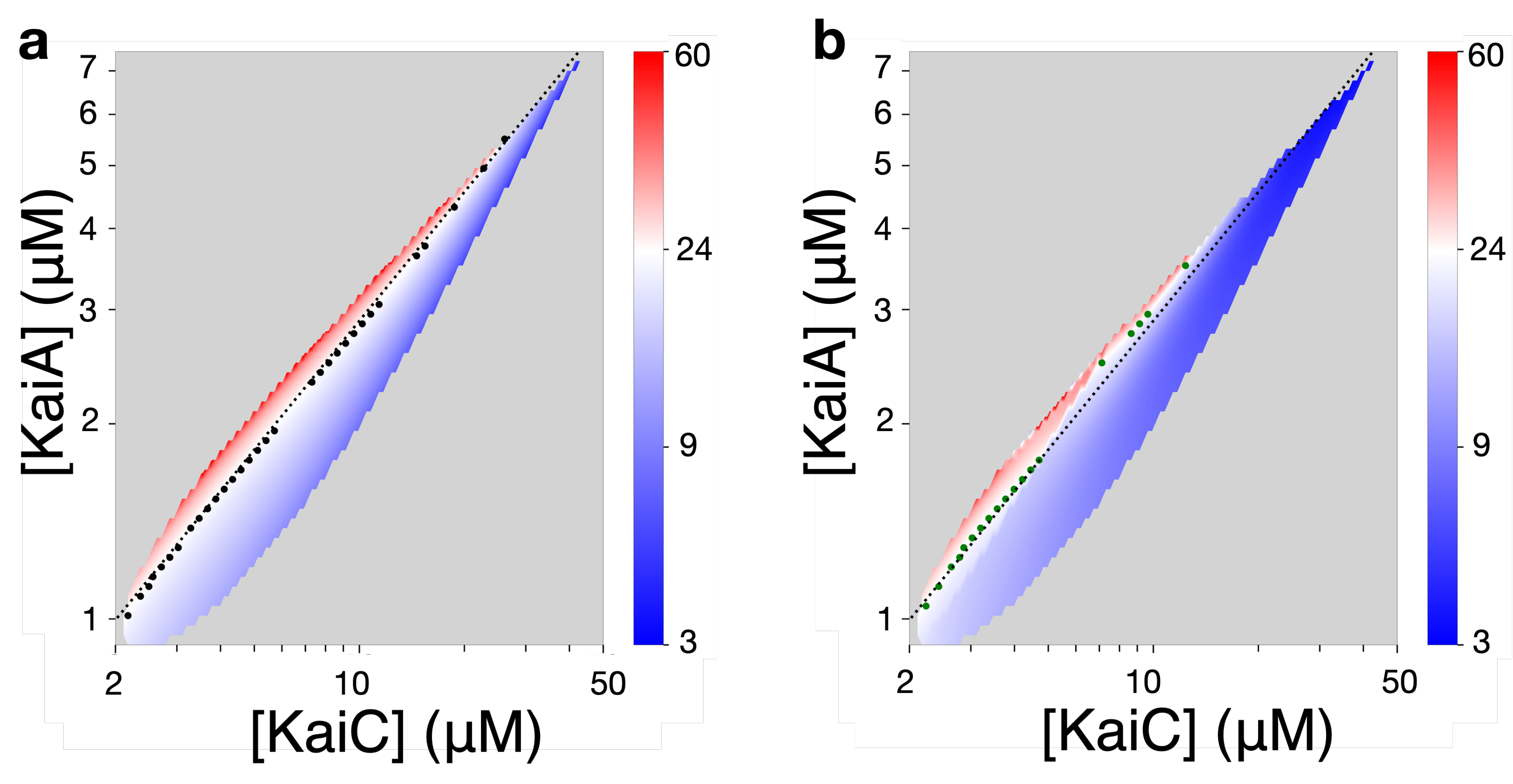}
    \caption{Heat maps of the oscillation period in Phase I obtained (a) by solving the ODEs in Eq.~\eqref{eq:kaiabc_ode} and (b) from the Stuart-Landau equation. The data points of $\langle T_{\rm os}\rangle = 24$ hr (filled circles) are depicted on both of the maps.}
    \label{fig:scaling}
\end{figure*}

	\begin{figure*}[htbp]
		\includegraphics[width=0.8\linewidth]{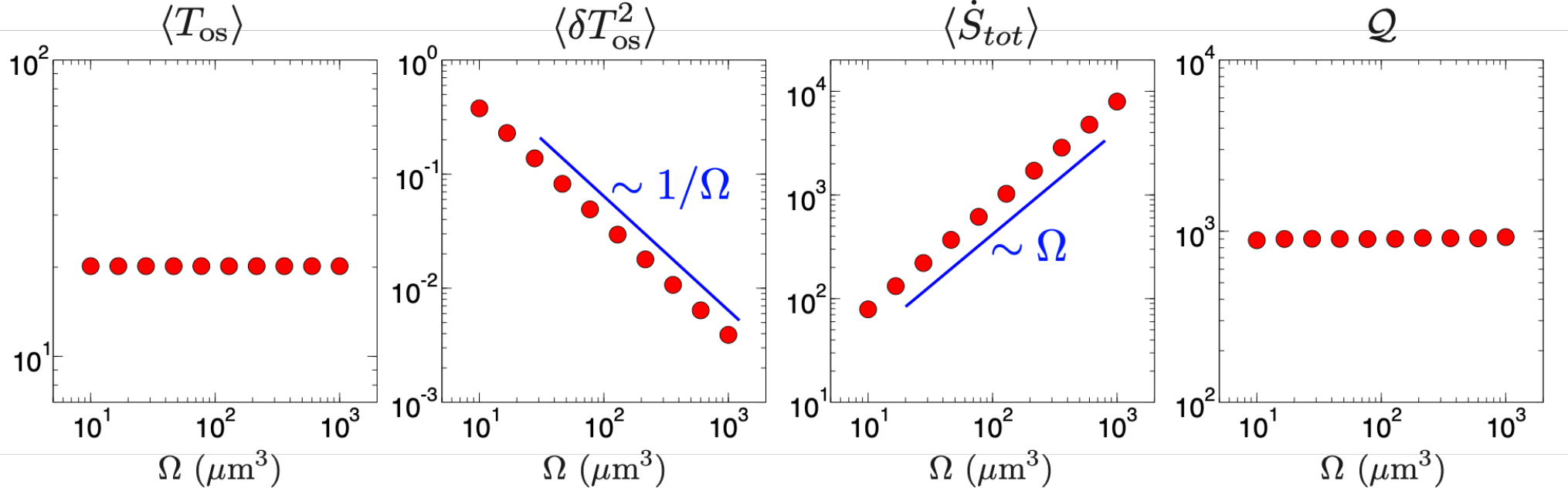}
		\caption{Period, variance, entropy production rate, and the uncertainty product of the KaiABC system with varying system size.}
		\label{fig:TUR_Omega}
		\end{figure*}
		
	\clearpage 
		
\begin{table}[h]
			\centering
			\renewcommand{\arraystretch}{1.1} 
			\setlength{\tabcolsep}{4pt} 
			\caption{Reaction rates and binding constants}
			\begin{tabular}{|c|c|}
				\hline
				$k_{UT}^{0}$ & 0 h$^{-1}$ \\
				$k_{TU}^{0}$ & 0.21 h$^{-1}$  \\
				\hline
				$k_{TD}^{0}$ & 0 h$^{-1}$ \\
				$k_{DT}^{0}$ & 0 h$^{-1}$  \\
				\hline
				$k_{DS}^{0}$ & 0.31 h$^{-1}$ \\
				$k_{SD}^{0}$ & 0 h$^{-1}$ \\
				\hline
				$k_{SU}^{0}$ & 0.11 h$^{-1}$ \\
				$k_{US}^{0}$ & 0 h$^{-1}$ \\
				\hline\hline
				$k_{UT}^{A}$ & 0.479077 h$^{-1}$ \\
				$k_{TU}^{A}$ & 0.0798462 h$^{-1}$ \\
				\hline
				$k_{TD}^{A}$ & 0.212923 h$^{-1}$ \\
				$k_{DT}^{A}$ & 0.1730000 h$^{-1}$ \\
				\hline
				$k_{DS}^{A}$ & $-0.319385$ h$^{-1}$ \\
				$k_{SD}^{A}$ & $0.505692$ h$^{-1}$ \\
				\hline
				$k_{SU}^{A}$ & $-0.133077$ h$^{-1}$ \\
				$k_{US}^{A}$ & 0.0532308 h$^{-1}$ \\
				\hline\hline
				{Binding const. KaiA-KaiC} & $K_{1/2}$ = 0.43 $\mu$M \\
				\hline
			\end{tabular}
			\label{table}
		\end{table}

\end{document}